\documentclass[10pt,english,aps,notitlepage,twocolumn,superscriptaddress,shortbibliography,nofootinbib,prl,floatfix]{revtex4-2}
\usepackage{mathtools,amsthm,amsmath,amsfonts,amssymb}
\usepackage{graphicx}
\usepackage{lipsum}
\usepackage{subfigure}
\usepackage[colorlinks = true,
            linkcolor = blue,
            urlcolor  = blue,
            citecolor = blue,
            anchorcolor = blue]{hyperref}
\usepackage{physics}
\usepackage{comment}
\usepackage{pgfplots}
    \usetikzlibrary{positioning}
\usepackage[normalem]{ulem}
\usepackage[noend]{algpseudocode}
\usepackage[dvipsnames]{xcolor}
\usepackage{orcidlink}
\usepackage{tikz}
    \usetikzlibrary{math}

\DeclareMathOperator{\mS}{\mathcal{S}}
\DeclareMathOperator{\sJ}{\mathcal{J}}
\def\hq{h^{(q)} }
\def\be{\begin{equation}}
\def\ee{\end{equation}}
\def\tcoh{t_{\text{coh}}}
\def\X{\mathbb X}
\def\Y{\mathbb Y}
\def\Z{\mathbb Z}

\newcommand{\sect}[1]{\emph{#1}---}

\newcommand{\ictpsaifradd}{ICTP South American Institute for Fundamental Research \\
                           Instituto de F\'{i}sica Te\'{o}rica, UNESP - Univ. Estadual Paulista \\
                           Rua Dr. Bento Teobaldo Ferraz 271, 01140-070, S\~{a}o Paulo, SP, Brazil}

\begin{document}
\title{{
Quantum Many-Body Scars from Operator Size\\
}}

\author{Debarghya Chakraborty \,\orcidlink{0009-0000-1078-5348}}
    \email{debarghya.chakraborty@ictp-saifr.org}
\affiliation{\ictpsaifradd}

\author{Dario Rosa\,\orcidlink{0000-0001-9747-1033}}
    \email{dario\_rosa@ictp-saifr.org}
    \affiliation{\ictpsaifradd}
\begin{abstract}
We present a general construction of approximate many-body scar states in two copies of a system using a bilocal size interaction that couples one site to its partner on the other copy. For an uncoupled two-sided Hamiltonian whose terms have fixed size in an operator basis, we identify polynomially many Krylov basis states that it annihilates and that, in the thermodynamic limit, become concentrated on a single size sector. The size operator splits them into a tower of equally spaced levels that gives rise to coherent revivals characterized by collective oscillations of energy across the two copies. We provide a concrete realization of this construction in two copies of a general $q$-body nearest-neighbor spin chain, and the SYK model. 
\end{abstract}
\maketitle
\sect{Introduction}
Many collective physical phenomena arise from the interplay of competing interaction terms. In this Letter, we present an example that leads to highly coherent non-ergodic dynamics between two identical copies of a quantum many-body system. The mechanism builds on \cite{Chakraborty:2025bhk} where it was observed that a Krylov number operator, when used as a coupling between the two systems, splits the exponentially large degenerate subspace of the uncoupled systems into a tower of equally spaced levels, \textit{i.e.} quantum many-body scars \cite{turner2018weak, turner2018quantum, schecter2019weak, iadecola2020quantum,bull2020quantum,serbyn2021quantum,wildeboer2022quantum, moudgalya2022quantum, chandran2023quantum, odea2025entanglement}. Within that subspace, the number operator acts like an emergent transverse field, giving rise to a collective coherent exchange of energy between the left and right copies of the system. Although completely general, the construction of \cite{Chakraborty:2025bhk} is algebraic in nature and, as such, it does not offer an explicit prescription for the form of the interaction operator in terms of the microscopic degrees of freedom. 

The specific example in \cite{Chakraborty:2025bhk} used two copies of the SYK model coupled by a fermion bilinear which measures the size of excitations on the infinite temperature thermofield double. Here we show that this correspondence is much more general: for a wide class of local spin systems, the Krylov basis states of fixed index built on the infinite-temperature thermofield double become size eigenstates in the thermodynamic limit. Consequently, the mechanism of \cite{Chakraborty:2025bhk} can be given a fully concrete implementation for these systems using a bilocal interaction that measures size. The setup can be implemented in state-of-the-art experimental platforms \cite{zhang2022many-body, su2023observation} and allows considerable freedom in the choice of the single-copy Hamiltonian. 

The underlying physical mechanism is simple. In many local systems, suitably rescaled infinite-temperature moments of the Hamiltonian converge to those of a normal distribution. This is because such systems have a small fraction of non-commuting pairs of terms. On choosing local Hamiltonians of uniform size in an operator basis, the orthogonalized Krylov vectors inherit a well-defined size with corrections that are small in the thermodynamic limit. The uncoupled two-sided Hamiltonian annihilates the Krylov states, and the competing size interaction splits them into equally spaced levels, producing collective coherent oscillations of energy between the copies.

These results show that the framework of \cite{Chakraborty:2025bhk} can be realized approximately through a physically realizable interaction, in a large class of systems. This complements other general mechanisms for systematically embedding quantum many-body scars in otherwise thermalizing systems~\cite{shiraishi2017systematic, moudgalya2022hilbert,moudgalya2023from, moudgalya2024exhaustive, odea2020from, pakrouski2020many-body, pakrouski2021group}.

\sect{Setup}
For the sake of concreteness, we will mostly focus on spin chains with open boundary conditions, written in terms of Pauli matrices $\sigma^{\alpha}_{k}$, with $\alpha \in \{x, y, z\}$ and $k \in \left\{ 1,2, \dots, N\right\}$ denoting the site, and with Hamiltonian
\be H = \sum^{i=N-q+1}_{i=1} h^{(q)}_i,\quad h^{(q)}_i =  \sum_{\alpha} J_{i, \alpha} \prod^{k= i+q-1}_{k=i} \sigma^{\alpha}_{k}. \label{eq:one_copy_Hamiltonian} \ee
With the notation $h^{(q)}_i$, we mean a $q$-body interaction at one of the $N-q+1$ bonds labeled by the index of its first site $i$. We allow the couplings $J_{i, \alpha}$ to depend on $i$, but when referring to a homogeneous chain we use $J_{x}, J_{y}, J_z$. The variance of $h^{(q)}_i$ reads
\be \sum_{\alpha \in \{x, y, z\} } J^2_{i, \alpha} = \sJ^2_i, \quad \expval{\sJ^2_i} = \sJ^2,\ee 
where $\expval{\cdot}$ denotes an ensemble average, when applicable.

Our construction is built upon the infinite-temperature thermofield double: 
\be \ket{\Omega} = \frac{1}{\sqrt{2^{N}}} \sum^{2^N}_{n=1} \ket{E_n}_{1} \otimes \Theta \ket{E_n}_{2}. \label{eq:def_infinite_tfd} \ee 
The antiunitary $\Theta$ implements the isomorphism between the operator space $\mathcal{B}(\mathcal{H})$ acting on a single copy and the doubled Hilbert space $\mathcal{H}\otimes\bar{\mathcal{H}}$ \cite{Choi1975, Jamiolkowski1972, Lensky_2021}, which we use extensively. We use $1$ and $2$ to denote the two copies of the system, and $\{\ket{E_n} \}$ denotes the energy eigenstates of $H$. We define
\be H_L := H_1 \otimes \mathbb I_2, \quad H_{R} := \mathbb I_1 \otimes (\Theta H \Theta^{-1} )_{2},  \ee
implying
\be (H_L - H_R) \ket{\Omega} = 0. \label{eq:hl_minus_hr_tfd} \ee
For a general operator $A$, we define
\be A_L := A_1 \otimes \mathbb I_2, \quad A_R := \mathbb I_1 \otimes  (\Theta A^{\dagger} \Theta^{-1})_2, \label{eq:def_A_L} \ee
which, through the isomorphism, defines the state 
\be \ket{A} := A_L \ket{\Omega} = A_R \ket{\Omega}. \label{eq:def_ket_A}\ee
Eq.~\eqref{eq:def_ket_A} shows that $\ket{\Omega}$ corresponds to $A_1 = \mathbb I$. Eqs.~\eqref{eq:def_ket_A}, \eqref{eq:def_A_L} can be used to define the one-sided operator $A_1$ from any two-sided state $\ket{A}$. This map turns an operator $O$, acting as $ O\ket{A} = \ket{B} $ on the two-copy system, into a superoperator $\mathcal{O}(\cdot)$ that acts as $\mathcal{O}(A) = B$. For instance, Eq.~\eqref{eq:hl_minus_hr_tfd} implies
\be (H_L - H_R) \ket{A} = (H_L A_L - A_L H_R ) \ket{\Omega} = \ket{[H, A]}, \ee
showing that $(H_L - H_R)$ becomes the Liouville superoperator $[H, \cdot]$. Eq.~\eqref{eq:hl_minus_hr_tfd} is simply the statement that $[H, \mathbb I]=0$. The space of operators on $N$ qubits can be expanded in the Pauli basis indexed by a set $I$ labeling the non-identity Paulis, 
\be A = \sum_{I} c_{I} P_{I}, \label{eq:pauli_decomp}\ee 
with a similar expansion for the state $\ket{A} = \sum_{I} c_{I} \ket{P_I}$. There are $3^{s} \binom{N}{s}$ strings of size $|I|=s$, which can be used to grade the Pauli basis. We introduce the term \textit{size distribution} to denote the probability distribution defined by $\sum_{\abs{I}=s} \abs{c_{I}}^2$ for a normalized $\ket{A}$. Its moments are given by expectation values of powers of the \textit{size operator} \cite{Qi_2019}
\be S = \sum_i S_i, \quad S_{i} = \frac{3}{4} - \frac{1}{4} \big( \sigma^{x}_{i, L} \sigma^{x}_{i, R}+ \sigma^{y}_{i, L} \sigma^{y}_{i, R} + \sigma^{z}_{i, L} \sigma^{z}_{i, R} \big). \label{eq:size_op_def} \ee
For a normalized $\ket{A}$, if $c_{I} \neq 0$ only for $\abs{I} = s$, then $\bra{A} S^{k} \ket{A} = s^k$. 

The inner product $\bra{A}\ket{B}$ becomes the Hilbert--Schmidt inner product $\tr(A^{\dagger} B)$, with $\tr$ being the normalized trace $\tr(\mathbb I)=1$. This makes the connection to the Hamiltonian moments $\mu_{k} := \tr(H^{k})$ clear, since  $\bra{H^{k_1}} \ket{H^{k_2}}=\mu_{k_1+k_2}$. We use $||A|| = \sqrt{\bra{A} \ket{A}} = \sqrt{\tr(A^{\dagger} A)}$.

\sect{Coupling the two copies}
We consider the coupled Hamiltonian (distinguished by a hat)
\be \hat{H}(\mu) = H_L - H_R + \mu S.  \label{eq:def_two_sided_Ham} \ee
For a factorized $\Theta$, in the sense of Sec.~\ref{sec:bell_state_basis}, $\sigma^{\alpha}_{i, R}$ is related by a local rotation to $\mathbb I_1 \otimes( \sigma^{\alpha}_{i} )_2$. Consequently $S_i$ couples the two spins on site $i$, one on each copy, and
\be \ket{\Omega} = \bigotimes^{i=N}_{i=1} \ket{\mathbb{I}}_i, \ee 
where $\ket{\mathbb{I}}_i$ is a maximally entangled state between the spins at the site $i$ on the two copies. The condition $S_i \ket{\mathbb{I}}_i = 0$ fixes $\ket{\mathbb{I}}_i$ up to a phase, in a basis-dependent form. Convenient choices for which $H_R = \pm \mathbb I\otimes H_2$ are discussed in Sec.~\ref{sec:bell_state_basis}.   

The Lanczos algorithm \cite{Lanczos:1950zz}, applied to $\ket{\Omega}$ with $H$, gives the normalized Krylov basis states $\ket{n}$. They satisfy $(H_L - H_R) \ket{n} =0$ and are defined by
\be \ket{n} := \frac{1}{\sqrt{r_n}} \ket{P_n(H)}, \ee
such that
\be \braket{P_m(H)}{P_n(H)} = r_n \delta_{mn}, \label{eq:orthogonality_krylov_vectors}  \ee
where $P_n(H)$ are orthogonal polynomials satisfying the three-term recursion of Eq.~\eqref{eq:recursion_for_monic_P}. $P_n(H)$ is defined to have leading coefficient $1$, \textit{i.e.}, $P_n(H)= H^{n} + \ldots$. 

Mathematically, our central result is that for fixed $n$ and $N \gg 1$, $\ket{n}$ gets concentrated on its maximal size sector. Concretely, we define 
\be \ket{D_n} := \mathcal{P}_{nq}  \ket{P_{n}(H)} =: D_n \ket{\Omega}, \label{eq:def_Dn} \ee  
with $\mathcal{P}_{nq}$ being the projector onto the maximal-size $(n q)$ sector. $D_n$ is the sum of products of $n$ energy density excitations $\hq$ -- with mutually disjoint spatial support
\be 
\label{eq:D_n_def}
D_n = n! \sum_{\substack{1 \leq i_1 < \ldots < i_n \leq N-q+1 \\ i_{k+1} - i_{k}  \geq q}} \hq_{i_1} \hq_{i_2} \ldots \hq_{i_n}.  \ee 
The state $\ket{1}$ is exactly $\frac{1}{\sqrt{\mu_2}} \ket{D_1}$.   
$\ket{D_n}$ can also be explicitly written in the spin basis as a sum over all possible $q$-mer coverings of the spin ladder with weights inherited from the Hamiltonian, as elaborated in Sec.~\ref{sec:bell_state_basis}.

In a decomposition analogous to Eq.~\eqref{eq:pauli_decomp}, the weight of $\ket{n}$ in the $nq$-sector is $\sum_{\abs{I} = nq} \abs{c_{I}}^2$. From Eq.~\eqref{eq:def_Dn} this quantity equals $||D_n||^2/r_n$. The numerator is simple to compute directly from the orthonormality of the Pauli basis:
\begin{subequations}
\begin{align}
||D_n||^2 &= (n!)^2 \sum_{\substack{1 \leq i_1 < \ldots < i_n \leq N-q+1 \\ i_{k+1} - i_{k}  \geq q}}  (\sJ_{i_1} \sJ_{i_2} \ldots \sJ_{i_n})^2, \\ 
||D_n||^2 &=  (n!)^2  \sJ^{2n} \binom{N - n  (q-1) }{n},\label{eq:avg_norm_dn}
\end{align}
\end{subequations}
where Eq.~\eqref{eq:avg_norm_dn} is exact for a homogeneous chain. For the disordered case we replace $\sJ^2_{i}$ by its average $\sJ^2$. This replacement is justified when $J_{i, \alpha}$ are independently distributed and $N$ is large. We will continue replacing $\sJ^2_i$ by its mean $\sJ^2$. When $n=1$, Eq.~\eqref{eq:avg_norm_dn} gives $\mu_2$.  

We define $\epsilon_n := 1- ||D_n||^2/r_n$ as the collective weight of $\ket{n}$ in all the size sectors smaller than $n q$. The convergence of $\mu_k/\mu_2^{\frac{k}{2}}$ to the moments of a standard normal distribution \cite{Hartmann_2005,Keating_2014random, Keating_2015, Erd_s_2014} for fixed $k$ and $N \rightarrow \infty$ implies that $\epsilon_n$ goes to 0 for fixed $n$. The leading order moments are $\mu_{2k}/ \mu_2^{k} \sim \frac{(2 k)!}{k! 2^k}$ and assuming that corrections are $O(1/N)$, it follows from Eq.~\eqref{eq:r_n_from_kappa} that $r_n = n! \mu^{n}_2 \big(1+O(\frac{1}{N} ) \big)$. Plugging the leading piece of $r_n$ gives
\be \epsilon_n = 1 - \frac{ n!  \binom{N - n  (q-1)  }{n}}{ (N-(q-1))^n } = \frac{n (n-1) (2 q-1)}{2 N} + O(N^{-2}). \label{eq:epsilon_n_naive}\ee 
Eq.~\eqref{eq:epsilon_n_naive} shows the suppression of $\epsilon_n$ with $N$ but misses the $1/N$ contribution from $r_n$. In Sec.~\ref{sec:norms_hn_from_cumulants}, we provide a derivation of the convergence of the rescaled $\mu_k$ to those of a standard normal distribution and the leading $1/N$ correction to $r_n$ using the Toda equations of motion \cite{Toda1993,Yang_Ismail,Dymarsky_2020,Angelinos:2025drf, Takahashi:2025iol}. This derivation assumes that the energy cumulants scale extensively and uses the recursion in Eq.~\eqref{eq:recursive_eom_bn} following \cite{Witte_1999} to get
\be r_n = n! \mu^{n}_2 \big( 1  + \frac{n(n-1)}{4} \frac{\mu_2(\mu_4 - 3 \mu^2_2) - \mu^2_3}{\mu^3_2} +   O(N^{-2} ) \big),  \label{eq:r_n_in_terms_of_mu}\ee 
from which the model-specific $1/N$ correction follows
\be  \epsilon_n =  \frac{n (n-1)}{4} \bigg( \frac{ 2(2 q-1)}{ N}  + \frac{\mu_2(\mu_4 - 3 \mu^2_2) - \mu^2_3}{\mu^3_2} + O(N^{-2} )  \bigg), \label{eq:epsilon_leading_order} \ee
for which we provide a numerical check in Sec.~\ref{sec:evaluation_spin_chain}.

Another way to quantify the width of the size distribution of $\ket{n}$ is through the variance $\bra{n} S^2 \ket{n} - \bra{n} S \ket{n}^2$. Denoting  $\sum_{\abs{I}=s} \abs{c_{I}}^2 = \abs{c_s}^2$, we rewrite $\bra{n} S^k \ket{n}$ as $(nq)^k + \sum_{s<nq} \abs{c_s}^2 (s^k - (nq)^k)$. Plugging this representation into the size variance gives 
\be \sum_{s<nq}  \abs{c_s}^2 (s - nq)^2 -\bigg( \sum_{s<nq}  \abs{c_s}^2 (s - nq) \bigg)^2. \label{eq:variance_expansion} \ee 
Dropping the negative second term in \eqref{eq:variance_expansion} and using that $(s - nq)^2 \leq (nq)^2$ gives the upper bound 
\be \bra{n} S^2 \ket{n} - \bra{n} S \ket{n}^2 \leq (n q)^2 \epsilon_n. \label{eq:size_variance_bound} \ee 

\sect{Validity of Krylov Size correspondence}
Eq.~\eqref{eq:epsilon_leading_order} shows that when $N \rightarrow \infty$, $\ket{n}$ converges to a size eigenstate for $n \sim O(1)$, and that this description definitely breaks down when $\epsilon_n \sim O(1)$, which eventually happens for $n \sim O(\sqrt{N})$. To show that $\ket{n}$ has well-defined size for all $n \sim O(N^{\alpha})$, with $0<\alpha<\frac{1}{2}$, we need control over the higher-order terms in the cumulant expansion. Due to the recursion in Eq.~\eqref{eq:recursive_eom_bn}, explicit checks show that $\log(r_n) - \log(n!\mu^n_2)$ is an expansion of the form $\sum_{k \geq 1} (f_k r_k(n))/N^k$, with $r_k(n)$ being a (system-independent) polynomial in $n$ of degree $k$ \cite{debarghya_2026_euclidean}. The system-dependence comes through the cumulants, whose contribution has been rewritten as $f_k/N^k$ where $f_k$ is $O(1)$, if we assume extensive cumulants. This strongly suggests that for the systems considered, $\ket{n}$ are approximate size eigenstates for all $n \sim O(N^{\alpha})$ with $\alpha$ being arbitrarily close to $\frac{1}{2}$. 

\sect{Initial States for Coherent Revivals}
Given the above considerations, any $ \ket{\psi_{0}} = \sum_{n} c_n \ket{n}$ with $\abs{c_n}^2$, concentrated in a window $n \ll \sqrt{N}$, will show coherent revivals of frequency $\mu$ when evolved with Eq.~\eqref{eq:def_two_sided_Ham}. Indeed, Eq.~\eqref{eq:epsilon_leading_order} shows that $\ket{n}$ behave like sharply peaked resonances in the spectrum of Eq.~\eqref{eq:def_two_sided_Ham}, equally spaced in energy and with a $\mu$-dependent mean. The general analytic features of the non-ergodic dynamics derived in \cite{Chakraborty:2025bhk} are preserved, with $H_L$ and $i [S, H_L]$ obeying Heisenberg equations of motion, identical in structure to those of a harmonic oscillator's position and momentum operators.

These oscillations imply an effective energy transfer between the two copies. The amount of energy transferred can be inferred from the structure of $\ket{\psi_0}$. Let us write $\ket{\psi_{0}}=\ket{g(H)} = \sum_n c^{g}_n \ket{n}$, where $g(x)$ is a function of $H$. To guarantee revivals, $\abs{c^{g}_n}^2$ must decay sufficiently rapidly for $n \gtrsim O(\sqrt{N})$. Therefore, the states $\ket{\psi_0}$ must have low Krylov complexity $\sum_{n} \abs{c^{g}_n}^2 n$ \cite{Parker_2019, Balasubramanian_2022, Nandy_2025, rabinovici_review_complexity, jeong2026_review}. States with  $g(H) = e^{-zH}$ with $z \in \mathbb C$ are precluded, unless $z$ scales as a negative power of $N$ (see Sec.~\ref{sec:size_of_tfd}).

In fact, any state with finite $\bra{\psi_0} H_L\ket{\psi_{0}} /N$ as $N \rightarrow \infty$ limit must have large support on states with at least $n \sim O(\sqrt{N})$. This follows from a simple scaling estimate from the Lanczos coefficients. If $c_n$ is sharply peaked at some $n_{*}$, then $\bra{\psi_0} H_L\ket{\psi_{0}}$ can at most be $O(\sqrt{n_{*} N})$ to leading order. $n_{*}$ needs to be $O(\sqrt{N})$ to get finite energy density. Shifting the weight to smaller $n$ preserves the revivals at the expense of the total energy $\bra{\psi_0} H_L\ket{\psi_{0}}$ scaling subextensively. This argument shows that the amount of energy the two copies can persistently exchange via this mechanism is limited to $N^{\alpha}$ with $\alpha < 1$, with an obstruction emerging at $\alpha = 1$. Nevertheless for any $O(1)$ choice of $\mu$, the scars supporting the revivals have finite energy density above the ground state, much like in most many-body scar constructions. 

The preparation of states $\ket{\psi_{0}}$ with support on small $n$ is computationally efficient, as the states $\ket{D_n}$ show. Although $\ket{D_n}$ is a superposition of $O(N^{n})$ maximally entangled states, it is obtained by acting with $D_n$ on $\ket{\Omega}$. From Eq.~\eqref{eq:D_n_def}, following for instance \cite{misguich2026introduction}, it is seen that $D_n$ can be written as a matrix product operator whose bond dimension grows with $n$ but not with $N$. Hence $\ket{D_n}$, and their linear combinations can be prepared with $O(N)$ resources. For isotropic couplings, $\ket{D_n}$ is a Dicke state of the excitations introduced in Sec.~\ref{sec:polymer_excitations}.

\begin{figure*}[ht]
  \centering
  \includegraphics[width=0.46\textwidth]{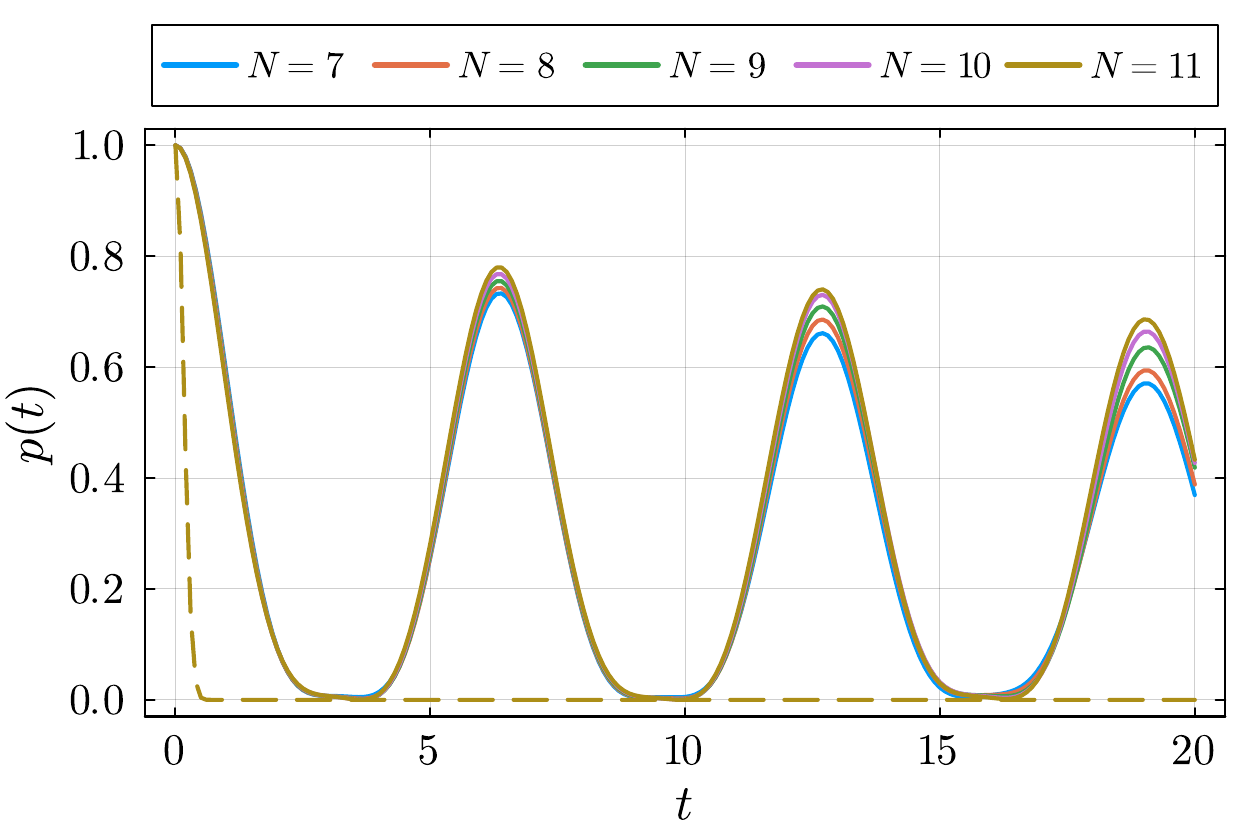}\hfill
  \includegraphics[width=0.46\textwidth]{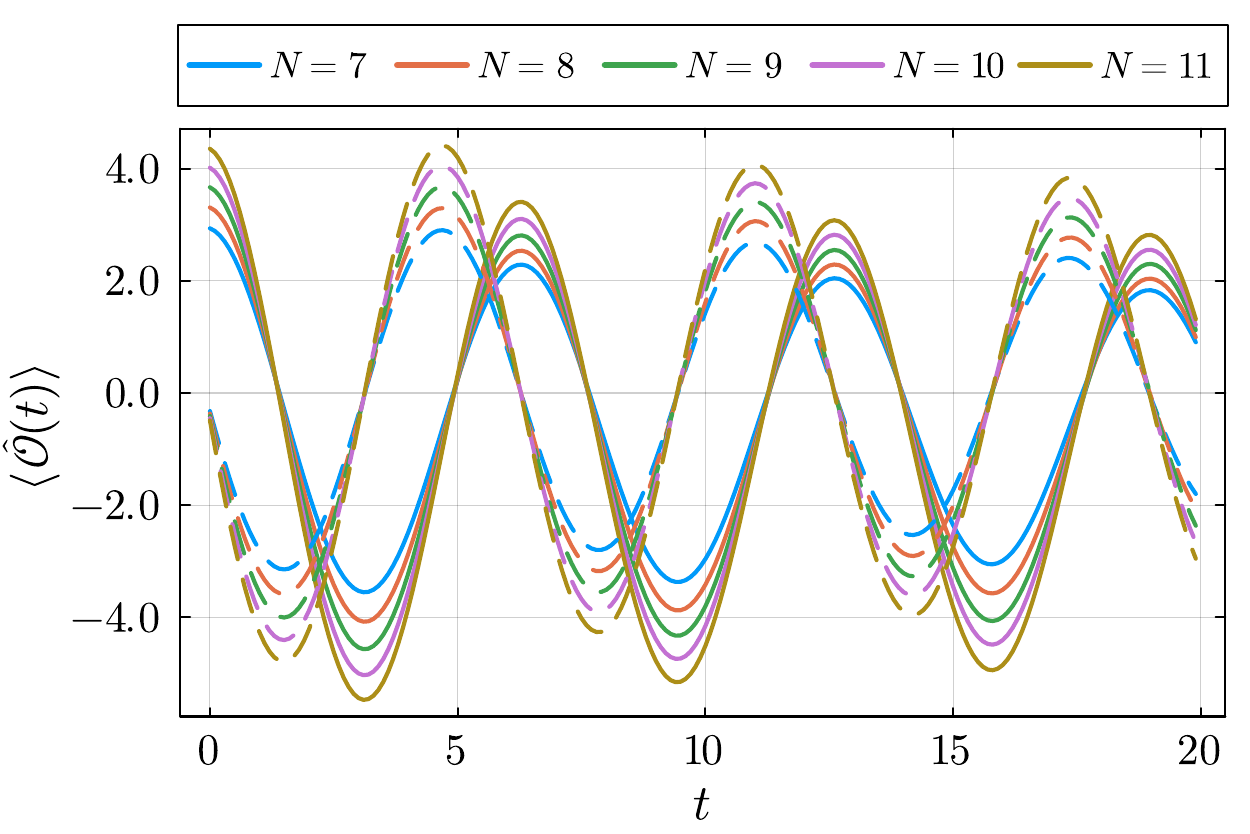}
   \caption{Time evolution for normalized $\ket{\psi_0} \propto \ket{g(H)}$ where $g(H)=\sum^{4}_{k=1} H^k$ across varying system sizes $N$. (a) Survival probability $p(t):=\abs{\bra{\psi_0} e^{-i\hat{H}(\mu) t} \ket{\psi_0}}^2$, which shows approximate revivals of period $2 \pi /(\mu q)$, whose quality improves with $N$. (b) Expectation values $\expval{\hat{\mathcal{O}}(t)}:= \bra{\psi_0} e^{i\hat{H}(\mu) t} \hat{\mathcal{O}} e^{-i\hat{H}(\mu) t} \ket{\psi_0}$, for $\hat{\mathcal{O}} = H_L$ (solid lines) and $\hat{\mathcal{O}}=i[S,H_L]$ (dashed lines), which show out-of-phase oscillations whose amplitude grows with $N$. $H$ is a $q=2$ homogeneous chain with $J_x=\frac{1}{\sqrt{3}}, J_y = \frac{1}{\sqrt{2}}, J_z= \frac{1}{\sqrt{6}}$ and $\mu=0.5$ in both panels.}
  \label{fig:revivals_inside_krylov}
\end{figure*}

\sect{Enhancement in SYK and coarse-grained size distribution}
The argument presented in the previous section does not \textit{rule out} revivals for high complexity states. Additional mechanisms are required to ensure that the action of $S$ on $\ket{\psi_0}$ does not cause it to leak appreciably outside of the Krylov space.

The SYK model in the double-scaling limit \cite{Berkooz_2018, Berkooz_2019} is a good example, with the extra structure coming from the coarse-grained size distribution \cite{Chakraborty:2025bhk}. Recall that SYK is defined by a $q$-body Hamiltonian of $N$ Majorana fermions (see Eq.~\eqref{eq:syk_hamiltonian}), with the double-scaling limit obtained by sending $q \rightarrow \infty, N \rightarrow \infty$ with $\lambda := \frac{2 q^2}{N}$ held fixed. Assuming self-averaging, the Krylov states $\ket{n}$ can be viewed as embeddings of auxiliary open chord states into the physical Hilbert space of SYK \cite{Lin_2022}. In this case, size counts the number of non-identity Majoranas. The coarse-grained size distribution of $\ket{n}$ follows from the relation between chord number and size \cite{Lin_2022, Rabinovici:2023yex, Ambrosini:2024sre}. The mean size of $\ket{n}$ is smaller than $n q$ by an amount finite in $\lambda$, and likewise the size variance of $\ket{n}$ and $\epsilon_n$ are non-vanishing for nonzero $\lambda$. This is in contrast to what we find for fixed-$q$ spin chains where $\epsilon_n$ and, consequently,  the size variance goes to $0$ as $N \rightarrow \infty$. The resolution of the apparent paradox is offered by the spread of the size distribution of $\ket{n}$, which is negligible compared to the size separation of $\sim q$ between successive $\ket{n}$. As a result, an $O(1)$ value of $\epsilon_n$ does not conflict with the existence of revivals.

We repeat the combinatorial analysis performed for geometrically local spin chains to characterize the projection onto the size $nq$ sector of $\ket{n}$ in SYK and present it in Sec.~\ref{sec:size_in_SYK}. We use the results of \cite{Garc_a_Garc_a_2017, Garc_a_Garc_a_2018}, who computed the $\frac{1}{N}$ expansion of $\mu_k$ at fixed $q$, to obtain $r_n$ in SYK to leading order. Interestingly, the $1/N$ contributions cancel if the disorder average is performed separately for $||D_n||^2$ and the moments in $r_n$. Thus in SYK, $\epsilon_n = O(N^{-2})$ for fixed $q$ and $n$. Our results make partial progress towards understanding how the ``would-be'' chord states are embedded in the Krylov space for a fixed disorder realization. 

\sect{Moving away from exact subspace}
An accessible family of approximate scars is found by relaxing the exact condition $(H_L - H_R) \ket{\psi_0} = 0$. We define
\be Q_a := \sum_{j \in I_a} h^{(q)}_j, \quad  \tilde{Q}_a := Q_a / ||Q_a||, \label{eq:def_Qa}\ee
with $I_a$ being a contiguous block of size $N_a$. By using well-separated mutually commuting $Q_a$, we build a continuous family of states
\be \ket{\psi_{0}} = \bigotimes^{a=m}_{a=1} \bigg( \cos(\theta_a) \ket{0_a}+ e^{i \phi_a} \sin(\theta_a) \ket{1_a} \bigg) \bigotimes_{i \notin I_a } \ket{\mathbb I}_i, \label{eq:psi_0} \ee
defined in terms of  
\be \ket{0_a} := \prod_{ i \in I_a} \ket{\mathbb I}_i,\quad \ket{1_a} :=  \tilde{Q}_a \prod_{ i \in I_a} \ket{\mathbb I}_i, \label{eq:def_zeroa_onea} \ee 
where in Eq.~\eqref{eq:def_zeroa_onea}, $\tilde{Q}_a$ denotes the restriction of Eq.~\eqref{eq:def_Qa} to a subsystem. For a single block, the timescale over which evolution under Eq.~\eqref{eq:def_two_sided_Ham} can be approximated by that of the interaction $\mu S$ is set by $||[H, \tilde{Q}_a]||^{-1}$. Because only the boundary terms contribute to the commutator, $||[H, \tilde{Q}_a]|| \propto ||Q_a||^{-1}$, yielding a single-block coherence time of $O(\sqrt{N_a})$. By spacing the blocks sufficiently, the leakage measured by the commutator becomes additive. For generic $\theta_a$, the timescale over which coherent revivals persist scales at least as $\tcoh \sim \sqrt{N_a/m}$. 

This $\ket{\psi_{0}}$ is a sum over size sectors that are multiples of $q$. Its projection to the $n q$ sector approximates $\ket{D_n}$ when $n^2 \ll m$. Preparing Eq.~\eqref{eq:psi_0} is only as hard as preparing independent copies of $H \ket{\Omega}$. This can be accomplished generally using the techniques of \cite{Childs:2012gwh}. We provide evidence for these claims in Sec.~\ref{sec:manifold_states_with_revivals}. 

In terms of the one-copy system, $D_n$ is the dominant fixed-size projection of the conserved operator $P_n(H)$. By contrast, the $Q_a$'s form a set of commuting energy-density excitations of fixed size. These relax slowly because energy can only flow through the surface of the block, whose volume could be large. Eq.~\eqref{eq:psi_0} defines an interpolating family of approximately conserved operators of fixed size. The form of $\ket{\psi_0}$ can be vastly generalized not only by considering blocks of various size, but also by modifying the expansion of $Q_a$ in terms of $h_j$. By choosing smooth coefficients, $\tcoh$ can be made $O(N_a)$ following the ``slow operator'' construction of \cite{Kim_2015}.

\sect{Discussion}
We presented a general construction of approximate many-body scars in a two-copy setup, with a number of scarred states that grows polynomially with the system size. The construction uses two ingredients: a Hamiltonian that is a sum over operator strings of size exactly $q$ in an orthogonal operator basis, and the locality of the interactions. We exemplified the construction in detail for the spin-$\frac{1}{2}$ chain with nearest-neighbor $q$-body interactions, but we stress that the combinatorial origin of the convergence of $\ket{n}$ to its leading size projection is far more general. We have shown that the convergence is enhanced in the SYK model at fixed $q$. We have checked that this extends, at least in some special cases, to hypercubic lattices in higher spatial dimensions and to systems with larger local Hilbert-space dimension. We leave a systematic characterization of the scope of our results to future work. 

Along these lines, two further directions are natural. First, the construction may extend to multipartite setups, where one could study energy transfer protocols on configurations with nontrivial geometry. Also, it does not preclude the coexistence of other mechanisms of ergodicity breaking within a single copy, and the interplay between these could be fascinating.

Our results build on past work that uses entanglement, the size coupling and the state-operator mapping between two copies of a system to produce highly coherent dynamics that would be unexpected for generic systems \cite{Chakraborty:2025bhk, Gao_2017, maldacena2018eternaltraversablewormhole, Maldacena_2017, Lensky_2021, langlett_2022, Schuster_2022, Brown_2023,Lychkovskiy_2024,garcia-garcia2025lindblad, Aguilar-Gutierrez:2026nmd}. On a single copy, the $D_n$ are spatially delocalized products of local energy densities. Their near conservation is distinct from conventional hydrodynamic slowing and raises the question of whether such structured operators of fixed size can be accessed through local probes, and whether the dynamics of their small nonconserved components reveal information about energy transport beyond conventional hydrodynamics. It would be nice to fully explain our results in the framework developed in \cite{Moudgalya_2024}. 

If $\ket{n}$ could be truncated to a few dominant, simple size sectors, such as $\ket{D_n}$ at fixed $n$, the Krylov expansion of $e^{i(H_L+H_R)t}\ket{\Omega}$ would constrain observables, see Sec.~\ref{sec:size_of_tfd} for an example. This depends on whether the size distribution of $\ket{n}$ admits a $1/N$ expansion analogous to that of $r_n$. Such an expansion would also advance the program of \cite{Chakraborty:2026ssu, Chakraborty:2026ndo} relating Krylov complexity distribution to a survival amplitude and we leave it for future work. A more ambitious target is the entanglement dynamics of this evolution. The simple entanglement structure of $\ket{D_n}$ suggested by its matrix product representation makes the leading order promising. 

Specific models could be constructed whose low-order moments are tuned to cause cancellations in the $1/N$ expansion of $\epsilon_n$. That would open up a wider class of initial configurations for observing the persistent flow of energy across the two systems. 

\sect{Acknowledgments} We thank Sergio Aguilar-Gutierrez, Anatoly Dymarsky and Zlatko Papic for constructive discussions. D.~R. acknowledges FAPESP, for the ICTP-SAIFR grant 2021/14335-0 and the Young Investigator (Jovem Pesquisador) grant 2023/11832-9. D.~R. also acknowledges the Simons Foundation for the Targeted Grant to ICTP-SAIFR. D.~C. acknowledges FAPESP support through the grant 2024/13100-8.

\bibliography{scars}

@article{Chakraborty:2025bhk,
    author = "Chakraborty, Debarghya and Rosa, Dario",
    title = "{Revival Dynamics from Equilibrium States: Scars from Chords in SYK}",
    eprint = "2512.16836",
    archivePrefix = "arXiv",
    primaryClass = "cond-mat.str-el",
    doi = "10.21468/SciPostPhys.21.3.064",
    journal = "SciPost Phys.",
    volume = "21",
    pages = "064",
    year = "2026"
}

@article{Witte_1999,
   title={The Lanczos algorithm for extensive many-body systems in the thermodynamic limit},
   volume={40},
   ISSN={1089-7658},
   url={http://dx.doi.org/10.1063/1.533010},
   DOI={10.1063/1.533010},
   number={10},
   journal={Journal of Mathematical Physics},
   publisher={AIP Publishing},
   author={Witte, N. S. and Bessis, D.},
   year={1999},
   month=Oct, pages={4975–4994} }

@article{Hartmann_2005,
   title={Spectral Densities and Partition Functions of Modular Quantum ystems as Derived from a Central Limit Theorem},
   volume={119},
   ISSN={1572-9613},
   url={http://dx.doi.org/10.1007/s10955-004-4298-5},
   DOI={10.1007/s10955-004-4298-5},
   number={5-6},
   journal={Journal of Statistical Physics},
   publisher={Springer Science and Business Media LLC},
   author={Hartmann, Michael and Mahler, Günter and Hess, Ortwin},
   year={2005},
   month=June, pages={1139–1151} }

@article{Keating_2015,
   title={Spectra and Eigenstates of Spin Chain Hamiltonians},
   volume={338},
   ISSN={1432-0916},
   url={http://dx.doi.org/10.1007/s00220-015-2366-0},
   DOI={10.1007/s00220-015-2366-0},
   number={1},
   journal={Communications in Mathematical Physics},
   publisher={Springer Science and Business Media LLC},
   author={Keating, J. P. and Linden, N. and Wells, H. J.},
   year={2015},
   month=May, pages={81–102} }

@misc{bhattacharya2026quadratic,
      title={The quadratic growth of Krylov spread complexity in the BTZ black hole}, 
      author={Aranya Bhattacharya and Mario Flory and Michal P. Heller and Emiliano Rizza and Tim Schuhmann},
      year={2026},
      eprint={2608.09922},
      archivePrefix={arXiv},
      primaryClass={hep-th},
      url={https://arxiv.org/abs/2608.09922}, 
}

@article{Lin_2022,
   title={The bulk Hilbert space of double scaled SYK},
   volume={2022},
   ISSN={1029-8479},
   url={http://dx.doi.org/10.1007/JHEP11(2022)060},
   DOI={10.1007/jhep11(2022)060},
   number={11},
   journal={Journal of High Energy Physics},
   publisher={Springer Science and Business Media LLC},
   author={Lin, Henry W.},
   year={2022},
   month=Nov }

@article{Garc_a_Garc_a_2017,
   title={Analytical spectral density of the Sachdev-Ye-Kitaev model at finite 
N},
   volume={96},
   ISSN={2470-0029},
   url={http://dx.doi.org/10.1103/PhysRevD.96.066012},
   DOI={10.1103/physrevd.96.066012},
   number={6},
   journal={Physical Review D},
   publisher={American Physical Society (APS)},
   author={García-García, Antonio M. and Verbaarschot, Jacobus J. M.},
   year={2017},
   month=Sept }

@article{Garc_a_Garc_a_2018,
   title={Exact moments of the Sachdev-Ye-Kitaev model up to order 1/N2},
   volume={2018},
   ISSN={1029-8479},
   url={http://dx.doi.org/10.1007/JHEP04(2018)146},
   DOI={10.1007/jhep04(2018)146},
   number={4},
   journal={Journal of High Energy Physics},
   publisher={Springer Science and Business Media LLC},
   author={García-García, Antonio M. and Jia, Yiyang and Verbaarschot, Jacobus J. M.},
   year={2018},
   month=Apr }

@misc{debarghya_2026_euclidean,
      title={Work in Progress}, 
      author={Debarghya Chakraborty},
      year={2026},
}

@article{Childs:2012gwh,
    author = "Childs, Andrew M. and Wiebe, Nathan",
    title = "{Hamiltonian Simulation Using Linear Combinations of Unitary Operations}",
    eprint = "1202.5822",
    archivePrefix = "arXiv",
    primaryClass = "quant-ph",
    doi = "10.26421/QIC12.11-12-1",
    journal = "Quant. Inf. Comput.",
    volume = "12",
    number = "11&12",
    pages = "0901--0924",
    year = "2012"
}

@article{Schuster_2022,
   title={Many-Body Quantum Teleportation via Operator Spreading in the Traversable Wormhole Protocol},
   volume={12},
   ISSN={2160-3308},
   url={http://dx.doi.org/10.1103/PhysRevX.12.031013},
   DOI={10.1103/physrevx.12.031013},
   number={3},
   journal={Physical Review X},
   publisher={American Physical Society (APS)},
   author={Schuster, Thomas and Kobrin, Bryce and Gao, Ping and Cong, Iris and Khabiboulline, Emil T. and Linke, Norbert M. and Lukin, Mikhail D. and Monroe, Christopher and Yoshida, Beni and Yao, Norman Y.},
   year={2022},
   month=July }

@article{Brown_2023,
   title={Quantum Gravity in the Lab. I. Teleportation by Size and Traversable Wormholes},
   volume={4},
   ISSN={2691-3399},
   url={http://dx.doi.org/10.1103/PRXQuantum.4.010320},
   DOI={10.1103/prxquantum.4.010320},
   number={1},
   journal={PRX Quantum},
   publisher={American Physical Society (APS)},
   author={Brown, Adam R. and Gharibyan, Hrant and Leichenauer, Stefan and Lin, Henry W. and Nezami, Sepehr and Salton, Grant and Susskind, Leonard and Swingle, Brian and Walter, Michael},
   year={2023},
   month=Feb }

@article{Kim_2015,
   title={Slowest local operators in quantum spin chains},
   volume={92},
   ISSN={1550-2376},
   url={http://dx.doi.org/10.1103/PhysRevE.92.012128},
   DOI={10.1103/physreve.92.012128},
   number={1},
   journal={Physical Review E},
   publisher={American Physical Society (APS)},
   author={Kim, Hyungwon and Bañuls, Mari Carmen and Cirac, J. Ignacio and Hastings, Matthew B. and Huse, David A.},
   year={2015},
   month=July }

@article{Lensky_2021,
   title={Rescuing a black hole in the large-q coupled SYK model},
   volume={2021},
   ISSN={1029-8479},
   url={http://dx.doi.org/10.1007/JHEP04(2021)116},
   DOI={10.1007/jhep04(2021)116},
   number={4},
   journal={Journal of High Energy Physics},
   publisher={Springer Science and Business Media LLC},
   author={Lensky, Yuri D. and Qi, Xiao-Liang},
   year={2021},
   month=Apr }

@article{Choi1975,
  author    = {Man-Duen Choi},
  title     = {Completely positive linear maps on complex matrices},
  journal   = {Linear Algebra and its Applications},
  volume    = {10},
  number    = {3},
  pages     = {285--290},
  year      = {1975},
  doi       = {10.1016/0024-3795(75)90075-0},
  url       = {https://doi.org/10.1016/0024-3795(75)90075-0}
}

@article{Jamiolkowski1972,
  author    = {Andrzej Jamio{\l}kowski},
  title     = {Linear transformations which preserve trace and positive semidefiniteness of operators},
  journal   = {Reports on Mathematical Physics},
  volume    = {3},
  number    = {4},
  pages     = {275--278},
  year      = {1972},
  doi       = {10.1016/0034-4877(72)90011-0},
  url       = {https://doi.org}
}

@misc{gao2024_commuting_syk,
      title={Commuting SYK: a pseudo-holographic model}, 
      author={Ping Gao},
      year={2024},
      eprint={2306.14988},
      archivePrefix={arXiv},
      primaryClass={hep-th},
      url={https://arxiv.org/abs/2306.14988}, 
}

@article{Almheiri_2024,
   title={Quantum gravity of the Heisenberg algebra},
   volume={2024},
   ISSN={1029-8479},
   url={http://dx.doi.org/10.1007/JHEP08(2024)098},
   DOI={10.1007/jhep08(2024)098},
   number={8},
   journal={Journal of High Energy Physics},
   publisher={Springer Science and Business Media LLC},
   author={Almheiri, Ahmed and Goel, Akash and Hu, Xu-Yao},
   year={2024},
   month=Aug }

@misc{berkooz2024_commuting,
      title={A Path Integral for Chord Diagrams and Chaotic-Integrable Transitions in Double Scaled SYK}, 
      author={Micha Berkooz and Nadav Brukner and Yiyang Jia and Ohad Mamroud},
      year={2024},
      eprint={2403.05980},
      archivePrefix={arXiv},
      primaryClass={hep-th},
      url={https://arxiv.org/abs/2403.05980}, 
}

@article{Dymarsky_2020,
   title={Quantum chaos as delocalization in Krylov space},
   volume={102},
   ISSN={2469-9969},
   url={http://dx.doi.org/10.1103/PhysRevB.102.085137},
   DOI={10.1103/physrevb.102.085137},
   number={8},
   journal={Physical Review B},
   publisher={American Physical Society (APS)},
   author={Dymarsky, Anatoly and Gorsky, Alexander},
   year={2020},
   month=Aug }

@article{Angelinos:2025drf,
    author = "Angelinos, Nikolaos and Chakraborty, Debarghya and Dymarsky, Anatoly",
    title = "{Temperature dependence in Krylov space}",
    eprint = "2508.19233",
    archivePrefix = "arXiv",
    primaryClass = "hep-th",
    doi = "10.1088/1751-8121/ae7167",
    journal = "J. Phys. A",
    volume = "59",
    number = "23",
    pages = "235204",
    year = "2026"
}

@article{Toda1993,
  author  = {Toda, Morikazu},
  title   = {Vibration of a Chain with Nonlinear Interaction},
  journal = {Series in Pure Mathematics},
  year    = {1993},
  pages   = {97--102},
  doi     = {10.1142/9789814354332_0014}
}

@article{Yang_Ismail,
doi = {10.1088/0305-4470/30/19/006},
url = {https://doi.org/10.1088/0305-4470/30/19/006},
year = {1997},
month = {oct},
publisher = {},
volume = {30},
number = {19},
pages = {6633},
author = {Yang Chen and Mourad E H Ismail},
title = {Thermodynamic relations of the Hermitian matrix ensembles},
journal = {Journal of Physics A: Mathematical and General},
}

@article{Erd_s_2014,
   title={Phase Transition in the Density of States of Quantum Spin Glasses},
   volume={17},
   ISSN={1572-9656},
   url={http://dx.doi.org/10.1007/s11040-014-9164-3},
   DOI={10.1007/s11040-014-9164-3},
   number={3-4},
   journal={Mathematical Physics, Analysis and Geometry},
   publisher={Springer Science and Business Media LLC},
   author={Erdős, László and Schröder, Dominik},
   year={2014},
   month=Dec, pages={441–464} }

@misc{Keating_2014random,
    author = "Keating, J. P. and Linden, N. and Wells, H. J.",
    title = "{Random matrices and quantum spin chains}",
    eprint = "1403.1114",
    archivePrefix = "arXiv",
    primaryClass = "math-ph",
    month = "7",
    year = "2014"
}

@article{Parker_2019,
   title={A Universal Operator Growth Hypothesis},
   volume={9},
   ISSN={2160-3308},
   url={http://dx.doi.org/10.1103/PhysRevX.9.041017},
   DOI={10.1103/physrevx.9.041017},
   number={4},
   journal={Physical Review X},
   publisher={American Physical Society (APS)},
   author={Parker, Daniel E. and Cao, Xiangyu and Avdoshkin, Alexander and Scaffidi, Thomas and Altman, Ehud},
   year={2019},
   month=oct }

@article{langlett_2022,
  title = {Rainbow scars: From area to volume law},
  author = {Langlett, Christopher M. and Yang, Zhi-Cheng and Wildeboer, Julia and Gorshkov, Alexey V. and Iadecola, Thomas and Xu, Shenglong},
  journal = {Phys. Rev. B},
  volume = {105},
  issue = {6},
  pages = {L060301},
  numpages = {6},
  year = {2022},
  month = {Feb},
  publisher = {American Physical Society},
  doi = {10.1103/PhysRevB.105.L060301},
  url = {https://link.aps.org/doi/10.1103/PhysRevB.105.L060301}
}

@article{Qi_2019,
   title={Quantum epidemiology: operator growth, thermal effects, and SYK},
   volume={2019},
   ISSN={1029-8479},
   url={http://dx.doi.org/10.1007/JHEP08(2019)012},
   DOI={10.1007/jhep08(2019)012},
   number={8},
   journal={Journal of High Energy Physics},
   publisher={Springer Science and Business Media LLC},
   author={Qi, Xiao-Liang and Streicher, Alexandre},
   year={2019},
   month=aug }

@misc{maldacena2018eternaltraversablewormhole,
      title={Eternal traversable wormhole}, 
      author={Juan Maldacena and Xiao-Liang Qi},
      year={2018},
      eprint={1804.00491},
      archivePrefix={arXiv},
      primaryClass={hep-th},
      url={https://arxiv.org/abs/1804.00491}, 
}

@article{Maldacena_2017,
   title={Diving into traversable wormholes},
   volume={65},
   ISSN={1521-3978},
   url={http://dx.doi.org/10.1002/prop.201700034},
   DOI={10.1002/prop.201700034},
   number={5},
   journal={Fortschritte der Physik},
   publisher={Wiley},
   author={Maldacena, Juan and Stanford, Douglas and Yang, Zhenbin},
   year={2017},
   month=may }

@article{Rabinovici:2023yex,
    author = "Rabinovici, E. and S{\'a}nchez-Garrido, A. and Shir, R. and Sonner, J.",
    title = "{A bulk manifestation of Krylov complexity}",
    eprint = "2305.04355",
    archivePrefix = "arXiv",
    primaryClass = "hep-th",
    doi = "10.1007/JHEP08(2023)213",
    journal = "JHEP",
    volume = "08",
    pages = "213",
    year = "2023"
}

@article{Ambrosini:2024sre,
    author = "Ambrosini, Marco and Rabinovici, Eliezer and S{\'a}nchez-Garrido, Adri{\'a}n and Shir, Ruth and Sonner, Julian",
    title = "{Operator K-complexity in DSSYK: Krylov complexity equals bulk length}",
    eprint = "2412.15318",
    archivePrefix = "arXiv",
    primaryClass = "hep-th",
    reportNumber = "CERN-TH-2025-040",
    doi = "10.1007/JHEP08(2025)059",
    journal = "JHEP",
    volume = "08",
    pages = "059",
    year = "2025"
}

@misc{garcia-garcia2025lindblad,
      title={Lindblad many-body scars}, 
      author={Antonio M. García-García and Zhongling Lu and Lucas Sá and Jacobus J. M. Verbaarschot},
      year={2025},
      eprint={2503.06665},
      archivePrefix={arXiv},
      primaryClass={quant-ph},
      url={https://arxiv.org/abs/2503.06665}, 
}

@article{turner2018weak,
author={Turner, C. J.
and Michailidis, A. A.
and Abanin, D. A.
and Serbyn, M.
and Papi{\'{c}}, Z.},
title={Weak ergodicity breaking from quantum many-body scars},
journal={Nature Physics},
year={2018},
month={Jul},
day={01},
volume={14},
number={7},
pages={745-749},
issn={1745-2481},
doi={10.1038/s41567-018-0137-5},
url={https://doi.org/10.1038/s41567-018-0137-5}
}

@article{Lychkovskiy_2024,
  title = {Duality between open systems and closed bilayer systems: Thermofield double states as quantum many-body scars},
  author = {Teretenkov, Alexander and Lychkovskiy, Oleg},
  journal = {Phys. Rev. B},
  volume = {110},
  issue = {24},
  pages = {L241105},
  numpages = {8},
  year = {2024},
  month = {Dec},
  publisher = {American Physical Society},
  doi = {10.1103/PhysRevB.110.L241105},
  url = {https://link.aps.org/doi/10.1103/PhysRevB.110.L241105}
}

@article{Moudgalya_2024,
   title={Symmetries as Ground States of Local Superoperators: Hydrodynamic Implications},
   volume={5},
   ISSN={2691-3399},
   url={http://dx.doi.org/10.1103/PRXQuantum.5.040330},
   DOI={10.1103/prxquantum.5.040330},
   number={4},
   journal={PRX Quantum},
   publisher={American Physical Society (APS)},
   author={Moudgalya, Sanjay and Motrunich, Olexei I.},
   year={2024},
   month=Nov }

@misc{misguich2026introduction,
  title = {Introduction to matrix-product states and tensor networks},
  author = {Grégoire Misguich},
  year = {2026},
  eprint = {2606.24803},
  archiveprefix = {arXiv},
  primaryclass = {cond-mat.str-el},
  url = {https://arxiv.org/abs/2606.24803},
}

@article{turner2018quantum,
  author = {Turner, C. J. and Michailidis, A. A. and Abanin, D. A. and Serbyn, M. and Papić, Z.},
  title = {Quantum scarred eigenstates in a Rydberg atom chain: Entanglement, breakdown of thermalization, and stability to perturbations},
  journal = {Physical Review B},
  year = {2018},
  volume = {98},
  issn = {2469-9969},
  url = {http://dx.doi.org/10.1103/PhysRevB.98.155134},
  doi = {10.1103/physrevb.98.155134},
  number = {15},
  publisher = {American Physical Society (APS)},
  month = Oct,
}

@article{serbyn2021quantum,
  author = {Serbyn, Maksym and Abanin, Dmitry A. and Papić, Zlatko},
  title = {Quantum many-body scars and weak breaking of ergodicity},
  journal = {Nature Physics},
  year = {2021},
  volume = {17},
  issn = {1745-2481},
  url = {http://dx.doi.org/10.1038/s41567-021-01230-2},
  doi = {10.1038/s41567-021-01230-2},
  number = {6},
  publisher = {Springer Science and Business Media LLC},
  month = May,
  pages = {675–685},
}

@article{moudgalya2022quantum,
  author = {Moudgalya, Sanjay and Bernevig, B Andrei and Regnault, Nicolas},
  title = {Quantum many-body scars and Hilbert space fragmentation: a review of exact results},
  journal = {Reports on Progress in Physics},
  year = {2022},
  volume = {85},
  issn = {1361-6633},
  url = {http://dx.doi.org/10.1088/1361-6633/ac73a0},
  doi = {10.1088/1361-6633/ac73a0},
  number = {8},
  publisher = {IOP Publishing},
  month = July,
  pages = {086501},
}

@article{chandran2023quantum,
  author = {Chandran, Anushya and Iadecola, Thomas and Khemani, Vedika and Moessner, Roderich},
  title = {Quantum Many-Body Scars: A Quasiparticle Perspective},
  journal = {Annual Review of Condensed Matter Physics},
  year = {2023},
  volume = {14},
  issn = {1947-5462},
  url = {http://dx.doi.org/10.1146/annurev-conmatphys-031620-101617},
  doi = {10.1146/annurev-conmatphys-031620-101617},
  number = {1},
  publisher = {Annual Reviews},
  month = Mar,
  pages = {443–469},
}

@article{schecter2019weak,
  author = {Schecter, Michael and Iadecola, Thomas},
  title = {Weak Ergodicity Breaking and Quantum Many-Body Scars in Spin-1 $XY$ Magnets},
  journal = {Physical Review Letters},
  year = {2019},
  volume = {123},
  issn = {1079-7114},
  url = {http://dx.doi.org/10.1103/PhysRevLett.123.147201},
  doi = {10.1103/physrevlett.123.147201},
  number = {14},
  publisher = {American Physical Society (APS)},
  month = Oct,
}

@article{wildeboer2022quantum,
  author = {Wildeboer, Julia and Langlett, Christopher M. and Yang, Zhi-Cheng and Gorshkov, Alexey V. and Iadecola, Thomas and Xu, Shenglong},
  title = {Quantum many-body scars from Einstein-Podolsky-Rosen states in bilayer systems},
  journal = {Physical Review B},
  year = {2022},
  volume = {106},
  issn = {2469-9969},
  url = {http://dx.doi.org/10.1103/PhysRevB.106.205142},
  doi = {10.1103/physrevb.106.205142},
  number = {20},
  publisher = {American Physical Society (APS)},
  month = Nov,
}

@article{iadecola2020quantum,
  author = {Iadecola, Thomas and Schecter, Michael},
  title = {Quantum many-body scar states with emergent kinetic constraints and finite-entanglement revivals},
  journal = {Physical Review B},
  year = {2020},
  volume = {101},
  issn = {2469-9969},
  url = {http://dx.doi.org/10.1103/PhysRevB.101.024306},
  doi = {10.1103/physrevb.101.024306},
  number = {2},
  publisher = {American Physical Society (APS)},
  month = Jan,
}

@article{odea2025entanglement,
  author = {O’Dea, Nicholas and Sriram, Adithya},
  title = {Entanglement Oscillations from Many-Body Quantum Scars},
  journal = {Physical Review Letters},
  year = {2025},
  volume = {134},
  issn = {1079-7114},
  url = {http://dx.doi.org/10.1103/PhysRevLett.134.210402},
  doi = {10.1103/physrevlett.134.210402},
  number = {21},
  publisher = {American Physical Society (APS)},
  month = May,
}

@article{bull2020quantum,
  author = {Bull, Kieran and Desaules, Jean-Yves and Papić, Zlatko},
  title = {Quantum scars as embeddings of weakly broken Lie algebra representations},
  journal = {Physical Review B},
  year = {2020},
  volume = {101},
  issn = {2469-9969},
  url = {http://dx.doi.org/10.1103/PhysRevB.101.165139},
  doi = {10.1103/physrevb.101.165139},
  number = {16},
  publisher = {American Physical Society (APS)},
  month = Apr,
}

@article{shiraishi2017systematic,
  author = {Shiraishi, Naoto and Mori, Takashi},
  title = {Systematic Construction of Counterexamples to the Eigenstate Thermalization Hypothesis},
  journal = {Physical Review Letters},
  year = {2017},
  volume = {119},
  issn = {1079-7114},
  url = {http://dx.doi.org/10.1103/PhysRevLett.119.030601},
  doi = {10.1103/physrevlett.119.030601},
  number = {3},
  publisher = {American Physical Society (APS)},
  month = July,
}

@article{Balasubramanian_2022,
   title={Quantum chaos and the complexity of spread of states},
   volume={106},
   ISSN={2470-0029},
   url={http://dx.doi.org/10.1103/PhysRevD.106.046007},
   DOI={10.1103/physrevd.106.046007},
   number={4},
   journal={Physical Review D},
   publisher={American Physical Society (APS)},
   author={Balasubramanian, Vijay and Caputa, Pawel and Magan, Javier M. and Wu, Qingyue},
   year={2022},
   month=Aug }

@article{pakrouski2020many-body,
  author = {Pakrouski, K. and Pallegar, P. N. and Popov, F. K. and Klebanov, I. R.},
  title = {Many-Body Scars as a Group Invariant Sector of Hilbert Space},
  journal = {Physical Review Letters},
  year = {2020},
  volume = {125},
  issn = {1079-7114},
  url = {http://dx.doi.org/10.1103/PhysRevLett.125.230602},
  doi = {10.1103/physrevlett.125.230602},
  number = {23},
  publisher = {American Physical Society (APS)},
  month = Dec,
}

@article{pakrouski2021group,
  author = {Pakrouski, K. and Pallegar, P. N. and Popov, F. K. and Klebanov, I. R.},
  title = {Group theoretic approach to many-body scar states in fermionic lattice models},
  journal = {Physical Review Research},
  year = {2021},
  volume = {3},
  issn = {2643-1564},
  url = {http://dx.doi.org/10.1103/PhysRevResearch.3.043156},
  doi = {10.1103/physrevresearch.3.043156},
  number = {4},
  publisher = {American Physical Society (APS)},
  month = Dec,
}

@article{Berkooz_2019,
   title={Towards a full solution of the large N double-scaled SYK model},
   volume={2019},
   ISSN={1029-8479},
   url={http://dx.doi.org/10.1007/JHEP03(2019)079},
   DOI={10.1007/jhep03(2019)079},
   number={3},
   journal={Journal of High Energy Physics},
   publisher={Springer Science and Business Media LLC},
   author={Berkooz, Micha and Isachenkov, Mikhail and Narovlansky, Vladimir and Torrents, Genis},
   year={2019},
   month=Mar}

@article{Berkooz_2018,
   title={Chord diagrams, exact correlators in spin glasses and black hole bulk reconstruction},
   volume={2018},
   ISSN={1029-8479},
   url={http://dx.doi.org/10.1007/JHEP08(2018)192},
   DOI={10.1007/jhep08(2018)192},
   number={8},
   journal={Journal of High Energy Physics},
   publisher={Springer Science and Business Media LLC},
   author={Berkooz, Micha and Narayan, Prithvi and Simón, Joan},
   year={2018},
   month=Aug }

@article{odea2020from,
  author = {O’Dea, Nicholas and Burnell, Fiona and Chandran, Anushya and Khemani, Vedika},
  title = {From tunnels to towers: Quantum scars from Lie algebras and  $q$-deformed Lie algebras},
  journal = {Physical Review Research},
  year = {2020},
  volume = {2},
  issn = {2643-1564},
  url = {http://dx.doi.org/10.1103/PhysRevResearch.2.043305},
  doi = {10.1103/physrevresearch.2.043305},
  number = {4},
  publisher = {American Physical Society (APS)},
  month = Dec,
}

@article{moudgalya2023from,
  author = {Moudgalya, Sanjay and Motrunich, Olexei I.},
  title = {From symmetries to commutant algebras in standard Hamiltonians},
  journal = {Annals of Physics},
  year = {2023},
  volume = {455},
  issn = {0003-4916},
  url = {http://dx.doi.org/10.1016/j.aop.2023.169384},
  doi = {10.1016/j.aop.2023.169384},
  publisher = {Elsevier BV},
  month = Aug,
  pages = {169384},
}

@article{moudgalya2022hilbert,
  author = {Moudgalya, Sanjay and Motrunich, Olexei I.},
  title = {Hilbert Space Fragmentation and Commutant Algebras},
  journal = {Physical Review X},
  year = {2022},
  volume = {12},
  issn = {2160-3308},
  url = {http://dx.doi.org/10.1103/PhysRevX.12.011050},
  doi = {10.1103/physrevx.12.011050},
  number = {1},
  publisher = {American Physical Society (APS)},
  month = Mar,
}

@article{moudgalya2024exhaustive,
  author = {Moudgalya, Sanjay and Motrunich, Olexei I.},
  title = {Exhaustive Characterization of Quantum Many-Body Scars Using Commutant Algebras},
  journal = {Physical Review X},
  year = {2024},
  volume = {14},
  issn = {2160-3308},
  url = {http://dx.doi.org/10.1103/PhysRevX.14.041069},
  doi = {10.1103/physrevx.14.041069},
  number = {4},
  publisher = {American Physical Society (APS)},
  month = Dec,
}

@article{zhang2022many-body,
  author = {Zhang, Pengfei and Dong, Hang and Gao, Yu and Zhao, Liangtian and Hao, Jie and Desaules, Jean-Yves and Guo, Qiujiang and Chen, Jiachen and Deng, Jinfeng and Liu, Bobo and Ren, Wenhui and Yao, Yunyan and Zhang, Xu and Xu, Shibo and Wang, Ke and Jin, Feitong and Zhu, Xuhao and Zhang, Bing and Li, Hekang and Song, Chao and Wang, Zhen and Liu, Fangli and Papić, Zlatko and Ying, Lei and Wang, H. and Lai, Ying-Cheng},
  title = {Many-body Hilbert space scarring on a superconducting processor},
  journal = {Nature Physics},
  year = {2022},
  volume = {19},
  issn = {1745-2481},
  url = {http://dx.doi.org/10.1038/s41567-022-01784-9},
  doi = {10.1038/s41567-022-01784-9},
  number = {1},
  publisher = {Springer Science and Business Media LLC},
  month = Oct,
  pages = {120–125},
}

@article{su2023observation,
  author = {Su, Guo-Xian and Sun, Hui and Hudomal, Ana and Desaules, Jean-Yves and Zhou, Zhao-Yu and Yang, Bing and Halimeh, Jad C. and Yuan, Zhen-Sheng and Papić, Zlatko and Pan, Jian-Wei},
  title = {Observation of many-body scarring in a Bose-Hubbard quantum simulator},
  journal = {Physical Review Research},
  year = {2023},
  volume = {5},
  issn = {2643-1564},
  url = {http://dx.doi.org/10.1103/PhysRevResearch.5.023010},
  doi = {10.1103/physrevresearch.5.023010},
  number = {2},
  publisher = {American Physical Society (APS)},
  month = Apr,
}

@misc{Chakraborty:2026ssu,
    author = "Chakraborty, Debarghya",
    title = "{Krylov Complexity from Loschmidt Amplitude}",
    eprint = "2607.10921",
    archivePrefix = "arXiv",
    primaryClass = "hep-th",
    month = "7",
    year = "2026"
}

@misc{Chakraborty:2026ndo,
    author = "Chakraborty, Debarghya",
    title = "{Universality and Emergent Classicality in Krylov Space}",
    eprint = "2611.03188",
    archivePrefix = "arXiv",
    primaryClass = "hep-th",
    month = "11",
    year = "2026"
}

@article{Gao_2017,
   title={Traversable wormholes via a double trace deformation},
   volume={2017},
   ISSN={1029-8479},
   url={http://dx.doi.org/10.1007/JHEP12(2017)151},
   DOI={10.1007/jhep12(2017)151},
   number={12},
   journal={Journal of High Energy Physics},
   publisher={Springer Science and Business Media LLC},
   author={Gao, Ping and Jafferis, Daniel Louis and Wall, Aron C.},
   year={2017},
   month=Dec }

@article{Aguilar-Gutierrez:2026nmd,
    author = "Aguilar-Gutierrez, Sergio E. and Kukolj, Trivko and Seitz, Josef",
    title = "{q-Askey deformations of double-scaled SYK}",
    eprint = "2605.13956",
    archivePrefix = "arXiv",
    primaryClass = "hep-th",
    doi = "10.1007/JHEP09(2026)196",
    journal = "JHEP",
    volume = "09",
    pages = "196",
    year = "2026"
}

@article{Lanczos:1950zz,
    author = "Lanczos, Cornelius",
    title = "{An iteration method for the solution of the eigenvalue problem of linear differential and integral operators}",
    doi = "10.6028/jres.045.026",
    journal = "J. Res. Natl. Bur. Stand. B",
    volume = "45",
    pages = "255--282",
    year = "1950"
}

@misc{rabinovici_review_complexity,
      title={Krylov Complexity}, 
      author={Eliezer Rabinovici and Adrián Sánchez-Garrido and Ruth Shir and Julian Sonner},
      year={2025},
      eprint={2507.06286},
      archivePrefix={arXiv},
      primaryClass={hep-th},
      url={https://arxiv.org/abs/2507.06286}, 
}

@article{Nandy_2025,
   title={Quantum dynamics in Krylov space: Methods and applications},
   volume={1125-1128},
   ISSN={0370-1573},
   url={http://dx.doi.org/10.1016/j.physrep.2025.05.001},
   DOI={10.1016/j.physrep.2025.05.001},
   journal={Physics Reports},
   publisher={Elsevier BV},
   author={Nandy, Pratik and Matsoukas-Roubeas, Apollonas S. and Martínez-Azcona, Pablo and Dymarsky, Anatoly and del Campo, Adolfo},
   year={2025},
   month=June, pages={1–82} }

@article{Takahashi:2025iol,
    author = "Takahashi, Kazutaka and Nandy, Pratik and del Campo, Adolfo",
    title = "{Krylov complexity under Hamiltonian deformations and Toda flows}",
    eprint = "2510.19436",
    archivePrefix = "arXiv",
    primaryClass = "quant-ph",
    reportNumber = "RIKEN-iTHEMS-Report-25",
    doi = "10.1103/zt9g-scp5",
    journal = "Phys. Rev. B",
    volume = "113",
    number = "14",
    pages = "144312",
    year = "2026"
}

@misc{jeong2026_review,
      title={Quantum Chaos and Spread of States in Krylov Subspace: A Topical Review}, 
      author={Hyun-Sik Jeong and Juan F. Pedraza},
      year={2026},
      eprint={2608.26825},
      archivePrefix={arXiv},
      primaryClass={hep-th},
      url={https://arxiv.org/abs/2608.26825}, 
}
\clearpage
\newpage
\setcounter{secnumdepth}{3}
\section{End Matter}
\subsection{Bell State Basis} \label{sec:bell_state_basis}
We restrict to $\Theta$ that have the standard antiunitary factorization $\Theta=\prod_{i} U_{i} \mathcal{K}_z $, 
where $U_{i}$ is a local unitary on the site $i$ and $\mathcal{K}_z$ is complex conjugation in the $\sigma^z$ basis. The choice
\be \Theta = \bigg(\prod^{i=N}_{i=1} \sigma^{y}_{i} \bigg) \mathcal{K}_z. \label{eq:choice_Theta}\ee 
is especially convenient because for arbirary single copy $H$,  $H_R = (-1)^q \mathbb I \otimes H_2$. This can be seen using 
\be \sigma^{\alpha}_{i,R} = - \mathbb I_1  \otimes  (\sigma^{\alpha}_{i})_{2} \label{eq:Theta_conv_Pauli}\ee
The condition $ \big( \sigma^{\alpha}_{i,L} - \sigma^{\alpha}_{i,R}  \big) \ket{\mathbb I}_i = 0$ for all $\alpha$ implies that 
\be \ket{\Omega} = \prod^{N}_{i=1} \bigotimes \ket{\mathbb I}_{i} \quad \text{with} \quad \ket{\mathbb I}_{i} = \ket{\Psi^{-}}_i,  \label{eq:bell_decomposition_tfd}\ee 
 where $\ket{\Psi^{-}}_{i}$ is the spin singlet, expressed in terms of the Bell pair notation 
 \be \ket{\Phi^{\pm}}_{i}   = \frac{1}{\sqrt{2}} \bigg( \ket{0}_{i,1} \ket{0}_{i,2} \pm \ket{1}_{i,1} \ket{1}_{i,2}  \bigg),\ee
 \be \ket{\Psi^{\pm}}_{i}  = \frac{1}{\sqrt{2}} \bigg( \ket{0}_{i,1} \ket{1}_{i,2} \pm \ket{1}_{i,1} \ket{0}_{i,2}  \bigg), \ee 

Other choices of $\Theta$ related to Eq.~\eqref{eq:choice_Theta} by a local unitary, transform the Bell pairs  Eq.~\eqref{eq:bell_decomposition_tfd} by multiplication and the Paulis in Eq.~\eqref{eq:Theta_conv_Pauli} by conjugation. For example $\Theta \rightarrow \bigg(\prod^{i=N}_{i=1} \sigma^{y}_{i} \bigg) \Theta$ turns $\Theta$ into $\mathcal{K}_z$ and transforms $\ket{\Omega}$ to 
\be  \bigg( \mathbb I_1 \otimes \bigg(\prod^{i=N}_{i=1} \sigma^{y}_{i} \bigg)_2  \bigg) \prod^{N}_{i=1} \ket{\Psi^{-}}_i = (-i)^N \prod^{N}_{i=1} \ket{\Phi^{+}}. \ee 
From the conjugation $\sigma^{y}_{i} \sigma^{\alpha}_{i} \sigma^{y}_{i}$, 
\be \sigma^{x/z}_{i, R} = \mathbb I_1 \otimes ( \sigma^{x/z}_{i} )_2,~~~ \mathbb \sigma^{y}_{i, R} = -\mathbb I_1 \otimes ( \sigma^{y}_{i} )_2 \ee
\subsection{Size Excitations \label{sec:polymer_excitations}}
We denote excitations $\sigma^{\alpha}\ket{\mathbb I}$ for $\alpha ={x, y, z}$ by $\{ \ket{\X}, \ket{\Y}, \ket{\Z} \}$ respectively. They form an orthogonal entangled basis for the two qubits.  
In terms of these $ h^{(2)}_i \ket{\Omega}$ looks like
\be  \big( J_{i,x} \ket{\X}_{i}\ket{\X}_{i+1} + J_{i,y} \ket{\Y}_{i} \ket{\Y}_{i+1} + J_{i,z} \ket{\Z}_i\ket{\Z}_{i+1} \big), \label{eq:dimer_excitations} \ee
where we omitted the $\ket{\mathbb I}$ Bell pairs at sites $j \neq {i, i+1}$. Eq.~\eqref{eq:dimer_excitations} can be interpreted as a sum over three flavors of \textit{dimers} with weights $J_{i, \alpha}$. We illustrate two dimer configurations that contribute for a $N=6$ chain and $n=2$. The dimer flavors are color-coded and their weights in the sum are shown. 
\begin{center}
     \includegraphics[width=0.35\textwidth]{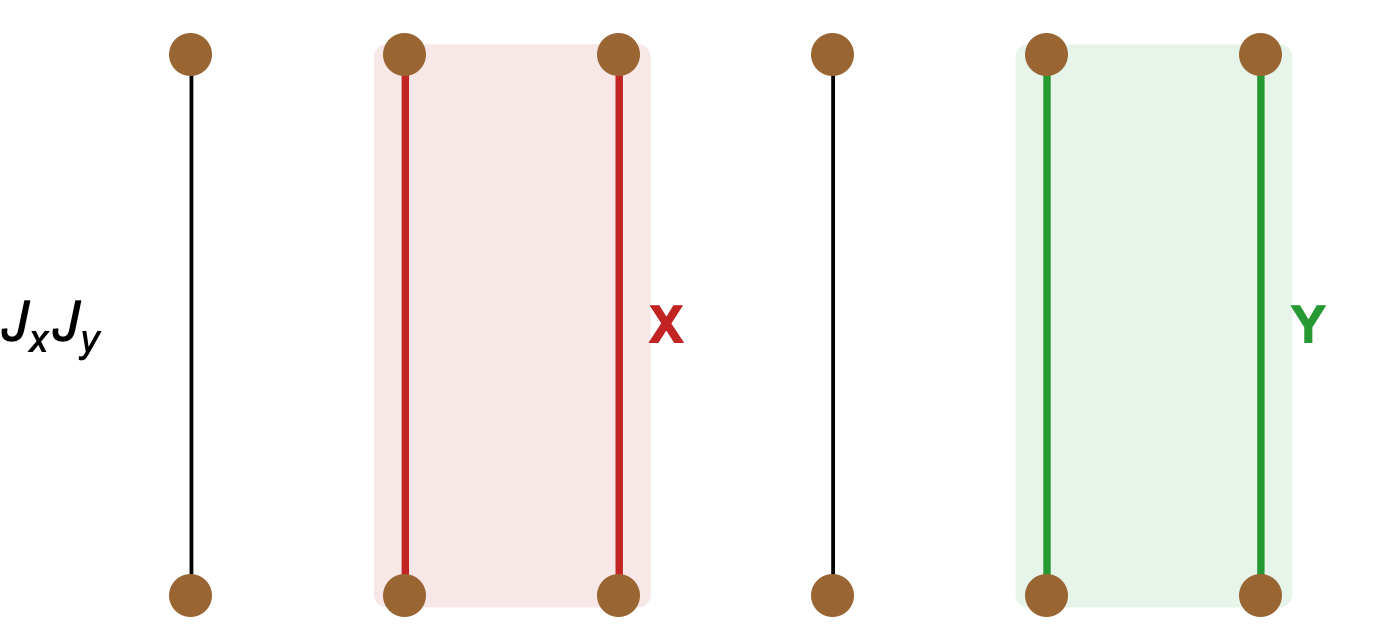}
\end{center}
\begin{center}
     \includegraphics[width=0.35\textwidth]{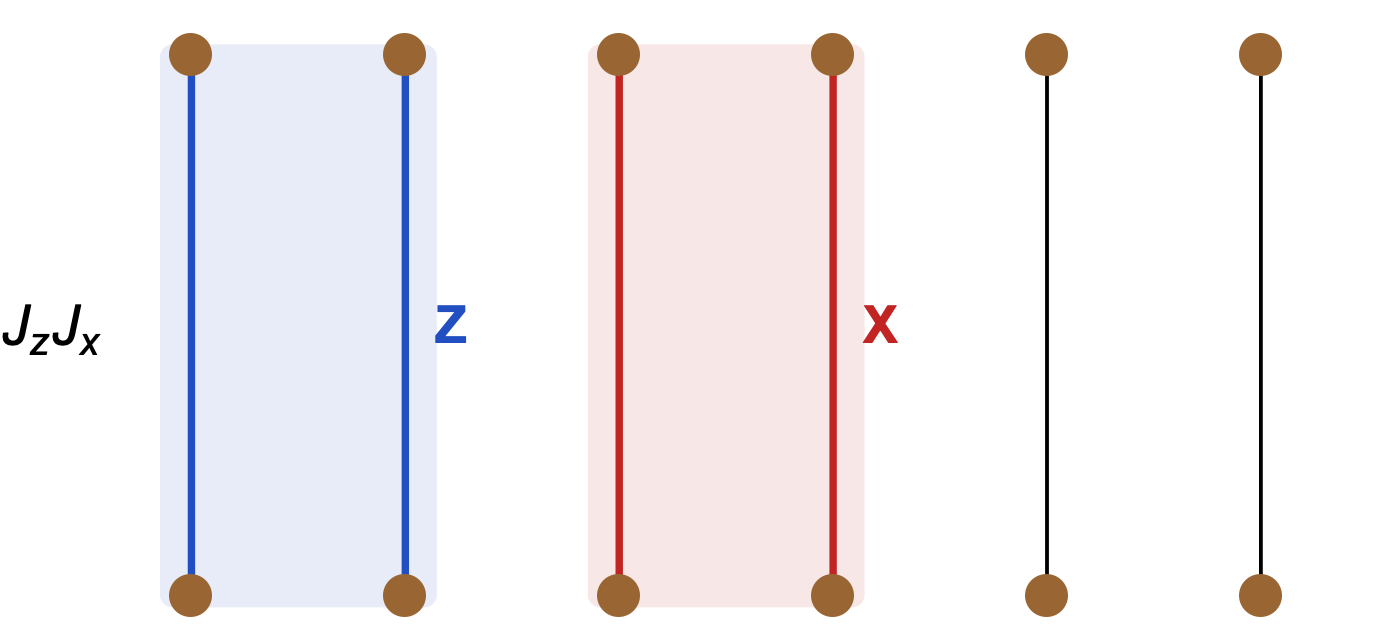}
\end{center}
The hardcore constraint is that any two dimers cannot overlap on the same Bell pair, since that would lower the size. Pictorially, two colors cannot fuse on a single rung of the ladder. The analogous expression for $h^{(3)}$ would instead involve a weighted sum over trimers, and so on for larger $q$. $\ket{D_n}$ can be interpreted as a sum over all possible configurations of exactly $n$ hardcore dimer ($q$-mer) excitations of $\ket{\Omega}$ with weights inherited from the Hamiltonian. When $J_x = J_y=J_z$, then all the configurations enter $\ket{D_n}$ with equal amplitude and the sum over color factorizes.

\subsection{The Krylov vectors}
\label{sec:Krylov_vectors}
The monic polynomials $P_n(H)$ satisfy 
\be  P_{n+1}(H) = (H-a_n) P_n(H) - b^2_{n-1} P_{n-1}(H), \label{eq:recursion_for_monic_P}  \ee 
in terms of the Lanczos coefficients $a_n$ and $b_n$, with boundary conditions $a_{-1}=0, b_{-1} = 0$. The Lanczos coefficients are determined by 
\be b^2_n = \frac{r_{n+1}}{r_n},~~~ a_n = \frac{1}{r_n} \bra{P_n(H)} H\ket{P_n(H)}.\ee 
The Lanczos coefficients are determined by the moments $\mu_k$. In terms of the density of states $\rho(E)$, the orthogonality relation Eq.~\eqref{eq:orthogonality_krylov_vectors} reads as 
\be  \int \rho(E) P_n(E) P_m(E) dE = r_n \delta_{nm}.\ee 
\subsection{Manifold of states with revivals}
\label{sec:manifold_states_with_revivals}
We take $q=2$ for specificity, omit the $q$-label on $h_j$ and consider $m$ contiguous blocks with $N_a$ sites and a buffer of $k$ sites between them. For example, we could take an interval defined by the indices $I_a = \{(a-1) (N_a +k) + 1, (a-1) (N_a +k) + 2, \ldots a (N_a +k) + (a-1)  k \}$ and use $Q_a$ as in Eq.~\eqref{eq:def_Qa}.

Then $\delta^2_{a} := ||(H_L - H_R) \ket{\tilde{Q}_a}||^2$ given by
\be   \frac{\tr([H, Q_a]^{\dagger} [H, Q_a]) }{\tr(Q^{\dagger}_a Q_a)} = \frac{||[h_{i_1-1} , h_{i_1} ||^2 +||[h_{i_{N_a}} , h_{i_{N_a}+1} ||^2 }{\sum_{i \in I_a} \sJ^2_{i}}, \label{eq:norm_of_suppressed} \ee
quantifies the magnitude of the component $\ket{\tilde{Q}_a}$ outside the Krylov space. $i_1$ and $i_{N_a}$ label the first and last index in $I_a$, and $h_{i_1-1}, h_{i_{N_a} +1 }$ are the terms in $H$ from outside $Q_a$. Clearly, $\delta^2_{a}$ is $O(1/N_{a})$ and can be made small if $N_a$ scales with $N$. 

$\delta_a$ sets the timescale for which time evolution by $H(\mu)$ can be approximated by time evolution with the free Hamiltonian $\mu S$. Applying the exact identity 
\be ( e^{- i \hat{H}(\mu) t} -  e^{-i \mu S t} \big)  = -i \int^{t}_{0} e^{-i \hat{H}(\mu) (t-x)} \big(H_L - H_R) e^{-i \mu S x} d x, \ee
to $\ket{\tilde{Q}_a}$ and applying the triangle inequality
\be || \big( e^{- i \hat{H}(\mu) t} -  e^{-i \mu S t} \big) \ket{\tilde{Q}_a}|| \leq \int^{t}_{0} ||\big(H_L - H_R) \tilde{Q}_a||  dx,  \label{eq:duhamel_eq} \ee 
along with the fact that $\tilde{Q}_a$ is a size eigenstate and that $e^{- i t\hat{H}(\mu)}$ is unitary. The right hand side of Eq.~\eqref{eq:duhamel_eq} evaluates to $\delta_a t$. This gives a conservative estimate for the revival dynamics with $t_{\text{coh}} = O(\frac{1}{\sqrt{N_a}})$

\begin{figure}
    \centering
    \includegraphics[width=0.95\linewidth]{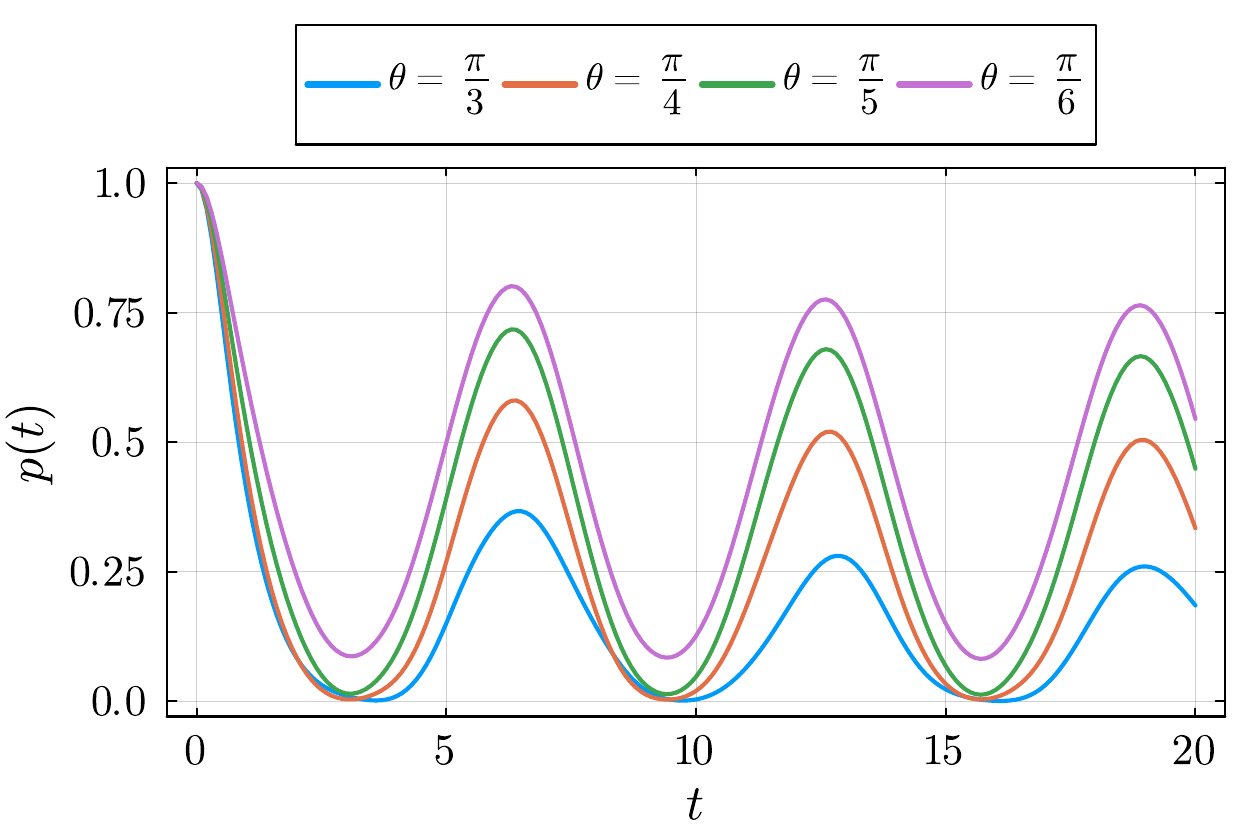}
    \caption{Survival probability $p(t)$ of the state $\ket{\psi_0}$ defined in Eq.~\eqref{eq:psi_0} for a chain of length $N=11$ where $m=2$ and $N_a=5$, with a buffer of one site. The angles $\theta_{1}=\theta_2 = \theta$ are shown in the legend.}
    \label{fig:factorized_psi0}
\end{figure}

The state in Eq.~\eqref{eq:psi_0} can be mapped to the single copy operator $\prod^{a=m}_{a=1} (\mathbb I \cos(\theta_a) + e^{i \phi_a} \sin(\theta_a) \tilde{Q}_a )$. Its commutator with $H$ simplifies to 
\be \sum^{a=m}_{a=1} e^{i \phi_a} \sin(\theta_a) [H, \tilde{Q}_a] \prod_{\substack{b \in \{1, \ldots m\} \\  b\neq a}} \big( \cos(\theta_b) + e^{i \phi_b} \sin(\theta_b) \tilde{Q}_b \big), \ee
where it is important for us to choose a buffer large enough that $[H, \tilde{Q}_a]$ commutes with each of the other $\tilde{Q}_b$. For $q=2$, this is ensured for $k \geq 1$. We use linearity and that traces factorize along the tensor product and that $||\big( \cos(\theta_b) + e^{i \phi_b} \sin(\theta_b) \tilde{Q}_b \big) || = 1$, to get 
\be
\begin{split}
\delta^2_{\psi_0}  := || (H_L - H_R) \ket{\psi_{0}}||^2  = \sum^{m}_{a=1} \sin^2(\theta_a) \frac{||[H, Q_{a}] ||^2}{|| Q_{a} ||^2}
\end{split}\ee 
The generalization to larger $q$ only requires modifying the number of buffer sites accordingly. If the $\sin(\theta_a)$ are $O(1)$ nonzero numbers, then $\delta_{\psi_0} \sim \sqrt{m/N_a}$. For $t \lesssim \tcoh$, 
\be e^{-i \hat{H}(\mu) t} \ket{\psi_{0}} \approx \bigotimes^{a=m}_{a=1} \bigg( \cos \theta_a \ket{0_a} + e^{i (\phi_a -  \mu q t)} \sin \theta_a \ket{1_a}  \bigg), \ee 
where we omitted the buffer Bell pairs that do not evolve. This implies the survival probability
\be \abs{\bra{\psi_0} e^{- i \hat{H}(\mu) t} \ket{\psi_0}}^2 \approx \prod^{a=m}_{a=1} \Bigg(1-\sin^{2}(2 \theta_a) \sin^2\bigg( \frac{1}{2} \mu q t \bigg) \Bigg),\ee 
where the rhs can be an useful upper bound when $N_a$ is not too large. For the same $t \lesssim t_{\text{coh}}$ regime when the leakage outside Krylov space is contained, $H_L$ and $i [S, H_L]$ continue to show out-of-phase oscillations. We provide an example in a finite $q=2$ chain in Fig.~\ref{fig:factorized_psi0} with the same homogeneous couplings as used as in Fig.~\ref{fig:revivals_inside_krylov}. 

The size $qn$ component of $\ket{\psi_0}$ has the form 
\be \sum_{i_1 i_2 \ldots i_n }  h_{i_1} h_{i_2} h_{i_n} \ket{\Omega}, \label{eq:prod_hn} \ee 
where we neglect overall normalization. In Eq.~\eqref{eq:prod_hn}, exactly $n$ of the $m$ disjoint blocks host a single excitation, and within each block all $N_a-q+1$ allowed positions of $h_i$ are equally likely. This gives $(N_a-q+1)^n\binom{m}{n}$ configurations, a subset of the $\binom{N-n(q-1)}{n}$ configurations in $\ket{D_n}$. For $N\approx mN_a \gg n^2 q$, $\binom{N-n(q-1)}{n}\approx N^n/n!$. Likewise, for $m\gg n^2$ and $N_a\gg nq$, $(N_a-q+1)^n\binom{m}{n}\approx (N_a m)^n/n!\approx N^n/n!$. 
The two counts agree to leading order. The only configurations of $\ket{D_n}$ absent from Eq.~\eqref{eq:prod_hn} are those with more than one excitation in a block, and these contribute at subleading order. The global superposition $\ket{D_n}$ (and thus $\ket{n}$ for fixed $n$) can therefore be mimicked by a stochastic product of excitations in separate blocks.


\clearpage
\setcounter{section}{0}
\setcounter{equation}{0}
\setcounter{figure}{0}
\setcounter{table}{0}
\renewcommand{\thesection}{SM \arabic{section}}
\renewcommand{\theequation}{S\arabic{equation}}
\renewcommand{\thefigure}{S\arabic{figure}}
\renewcommand{\thetable}{S\arabic{table}}
\setcounter{secnumdepth}{3}
\begin{center}\textbf{Supplemental Material}\end{center}
\section{Norms from extensive cumulants \label{sec:norms_hn_from_cumulants}}
The derivatives of the partition function 
\be Z(\beta) = \int dE \rho(E) e^{-\beta E},  \ee
give the $\beta$-dependent moments $\mu_{k}(\beta)$, where $\rho(E)$ is the density of states. Their values at $\beta=0$ determine the Lanczos coefficients and the norm $r_n$. It is convenient to consider their $\beta$-dependent generalization. These describe the data relevant for describing the Krylov basis constructed with the finite temperature thermofield double $e^{-\beta (H_L + H_R)/4 } \ket{\Omega}$ as the initial state for the Lanczos procedure. This was studied recently in \cite{bhattacharya2026quadratic}. The $\beta$-dependent trajectory of Lanczos coefficients, and their physical implications for Krylov complexity will be described in greater details in upcoming work by one of us \cite{debarghya_2026_euclidean}.   

The Hankel matrix of moments 
\be  (\mathbf{H}_{n}(\beta))_{ij} := \mu_{i+j}(\beta)~~~\quad~~~0 \leq i, j \leq n. \ee
encode all the relevant details through their determinant
\be  \tau_n(\beta) = \det \mathbf{H}_{n}(\beta).  \ee
In our conventions $\tau_0(\beta) = Z(\beta)$ and $\tau_{-1} = 1$. They determine
\begin{subequations}
\begin{align}
r_n(\beta) &= \frac{\tau_{n}(\beta)}{\tau_{n-1}(\beta)}, \label{eq:r_n_from_tau}   \\ 
b^2_n(\beta) &= \frac{r_{n+1}(\beta)}{r_n(\beta)},   \label{eq:bn2_in_terms_rn}\\
 a_n(\beta) &=  -\frac{d}{d\beta} \log r_n(\beta).     
 \end{align}
\end{subequations}
The equations of motion  Eq.~\eqref{eq:bn2_in_terms_rn} can be derived most simply with the substitution $\mathbf{q}_n = \log r_n$, and denotes the position of the $n$th Toda particle. In terms of these coordinates, $\tau_n = e^{\sum^{n}_{k=0} \mathbf{q}_k}$. The Toda equations imply 
\be b^2_n(\beta) = \frac{d^2 \log \tau_n}{d \beta^2}. \label{eq:log_hirota_toda} \ee 
Combining Eqs.~\eqref{eq:r_n_from_tau}, \eqref{eq:bn2_in_terms_rn} gives
\be \tau_n(\beta) = Z(\beta)^{n+1} \prod^{n}_{j=0} \big( b^2_j(\beta) \big)^{n-j}. \label{eq:tau_func_rec_def}  \ee 
\be  r_n(\beta) = Z(\beta) \prod^{n-1}_{j=0} b^2_j(\beta). \label{eq:rn_from_bn} \ee 
Plugging Eq.~\eqref{eq:tau_func_rec_def} into Eq.~\eqref{eq:log_hirota_toda} gives
\be  b^2_{n}(\beta) = (n+1) \frac{d^2}{d\beta^2} \log Z(\beta) + \sum^{n}_{j=0} (n-j) \frac{d^2}{d\beta^2}  \log(b^2_j). \label{eq:recursive_eom_bn}\ee 
We introduce $\mathcal{C}=b^2_0 = \frac{d^2}{d\beta^2} \log Z(\beta)$. Eq.~\eqref{eq:recursive_eom_bn} shows  that $b^2_n(\beta) = (n+1) \mathcal{C}(\beta) + \mathcal{C}^{0}(\beta)$ where $\mathcal{C}^{0}(\beta)$ is subleading in $N$ for fixed $n$. The derivatives of $\log Z(\beta)$ are assumed to be scale with an overall factor of $N$. Plugging in this ansatz into Eq.~\eqref{eq:recursive_eom_bn} yields
\be b^2_{n}(\beta) = (n+1) \mathcal{C}(\beta) + \sum^n_{j=0} (n-j) \frac{d^2}{d\beta^2} \log( \mathcal{C}(\beta) ) + O(\frac{1}{\mathcal{C}}),\ee
which we evaluate at $\beta=0$ 
\be b^2_{n} = (n+1) \kappa_2 +\frac{n (n+1)}{2} \frac{\kappa_2 \kappa_4 - \kappa^2_3}{\kappa^2_2} + O(N^{-2} ). \label{eq:b2n_from_kappa}  \ee 
in terms of the cumulants $\kappa_n$ defined by $\log Z(\beta) = \sum_{n\geq0}  \kappa_n \frac{\beta^{n}}{n!}$. We plug this into Eq.~\eqref{eq:rn_from_bn} at $\beta=0$ and find
\be r_n= n! \kappa^{n}_{2} \bigg(1+ \frac{n (n-1)}{4} \frac{\kappa_2 \kappa_4 - \kappa^2_3}{\kappa^3_2} + O(N^{-2}) \bigg). \label{eq:r_n_from_kappa} \ee 
These cumulants are expressed in terms of the moments as
\be \kappa_2 = \mu_2, \quad \kappa_3 = \mu_3, \quad \kappa_4 = \mu_4 - 3 \mu^2_2. \ee
which gives Eq.~\eqref{eq:r_n_in_terms_of_mu}, where used that $\mu_1=0$ for the systems we consider. The diagonal coefficients are 
\be a_n = - n \frac{\mu_3}{\mu_2} + O(N^{-1}). \ee 
If we define the symmetric tridiagonal matrix $M_{mn}= a_n \delta_{mn} + b_{n-1} \delta_{m,n-1}  + b_{n} \delta_{m, n+1}$, then the moments are given by 
\be \mu_{k} = (M^{k})_{00}. \ee 
It is seen that to leading order in large $N$, $b_n = \sqrt{n+1} \mu_2$ and $a_n =O(1)$. It follows from comparison with the Lanczos coefficients of the unperturbed Gaussian distribution that $\mu_{2k}/(\mu^{k}_2) =  \frac{(2k)!}{2^{k} k!}+ O(N^{-1})$ and $\mu_{2k+1}/(\mu_2)^{\frac{1}{2} + k} = O(N^{-1} )$  for $k\in \mathbb N$. The explicit form of the $1/N$ corrections are known from the Edgeworth expansion. 
\subsection{Evaluation in spin chain \label{sec:evaluation_spin_chain}}
\begin{figure}
    \centering
    \includegraphics[width=0.99\linewidth]{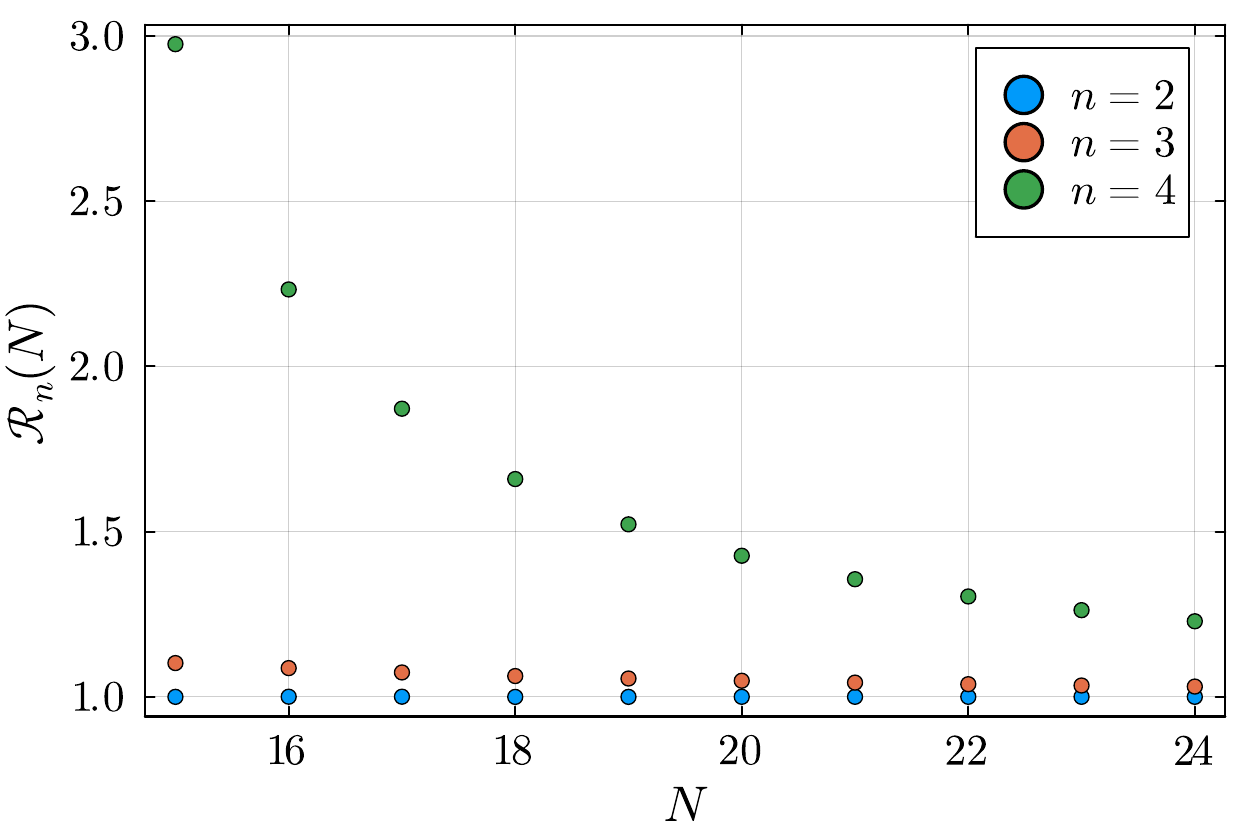}
    \caption{$\mathcal{R}_n(N):= ||D_n||^2/r_n$ plotted against $N$ for a homogeneous $q=2$ chain. Different colors correspond to $n$ shown in the legend. Here, $J_x=\frac{1}{\sqrt{3}}, J_y = \frac{1}{\sqrt{2}}, J_z= \frac{1}{\sqrt{6}}$.}
    \label{fig:ratios_plot}
\end{figure}

We evaluate the low order moments for a $q=2$ chain:
\be \mu_2 = \sum^{N-1}_{i=1} \sJ^2_i, \quad \mu_3 = -6 \sum^{N-1}_{i=1} J_{i, x} J_{i, y} J_{i, z}.  \ee 
\be  \mu_4 = 3 \mu^2_2 - 2 \sum^{N-1}_{i=1} \sum_{\alpha} J^4_{i,\alpha} - 4 \sum^{N-2}_{i=1} \sum_{\alpha \neq \beta} J^2_{i, \alpha} J^2_{i+1, \beta}. \ee
For the homogenous chain, these reduce to 
\be \mu_2 =  (N-1) \sJ^2, \quad \mu_3 = -6 (N-1) J_x J_y J_z, \ee 
\be
\begin{split}
    \mu_4 = 3 (N-1)^2 \sJ^4  - 
 2 (J_x^4 + J_y^4 + J_z^4) (N-1)   \\ 
 -8 (J_x^2 J^2_z  + J_y^2 J_z^2 + J_y^2 J_x^2 )(N -2),
\end{split}
\ee 
which gives rise to  
\be \epsilon_n = \frac{ n(n-1)\bigg( 2\sJ^2 (J^4_{x}+ J^4_y + J^4_z  + \sJ^4 ) - (6 J_{x} J_y J_z)^2   \bigg) }{ 4\sJ^6 N }, \label{eq:epsilon_n_clean_q=2} \ee
to leading order in $1/N$. We present an example of the convergence of $||D_n||^2/r_n \rightarrow 1$ with $N$ in Fig.~\ref{fig:ratios_plot}. 

\section{Size Distribution of SYK Krylov Vectors \label{sec:size_in_SYK}}
The SYK model describes $N$ Majorana fermions $\psi_i$ that have $q$-body interactions with couplings $J_{i_1, i_2, \ldots i_q}$ drawn from a Gaussian random distribution. The Majoranas are normalized as $\{ \psi_i, \psi_j \} = 2\delta_{ij}$. The Hamiltonian can be written compactly as a sum over $\binom{N}{q}$ ordered index sets $I = \{i_1, i_2, \ldots  i_q\}$ where $1 \leq i_1< i_2 <\ldots i_{q-1}<i_q \leq N$.     
\be H = i^{\frac{q}{2}}\sum_{I} J_{I} \Psi_{I},\quad \Psi_{I}= \prod_{i \in I} \psi_i, \label{eq:syk_hamiltonian} \ee
The ensemble averages over Gaussian couplings are given by
\be \expval{J_{I}} = 0, \quad \expval{J_{I} J_{K}} = \frac{\sJ^2}{ \lambda} \binom{N}{q}^{-1}  \delta_{I K}, \label{eq:SYK_ensemble_avg_conventions} \ee
in terms of $\lambda =2 q^2/N$. The ensemble averaged variance follows
\be \mu_2  = \frac{\sJ^{2}}{\lambda}. \ee 
The maximal size projection of $\mathcal{P}_{nq}\ket{P_n(H)}$ in SYK is defined by $D_n \ket{\Omega}$ where 
\be D_n = \sum_{I_1, I_2 \ldots I_n} J_{I_1} J_{I_2} \ldots J_{I_n} \Psi_{I_1} \Psi_{I_2} \ldots \Psi_{I_n} \label{eq:syk_Dn}\ee 
and there are exactly zero pairwise intersections among the index sets, i.e, $| I_{i} \cap I_k| = 0$. Since there are no geometric constraints it is easy to directly enumerate the cardinality of distinct terms entering the sum Eq.~\eqref{eq:syk_Dn} 
\be \binom{N}{q}  \binom{N-q}{q}  \ldots \binom{N-(n-1) q}{q} = \frac{N!}{(N-n q)! (q!)^n}.  \ee 
To compute the norm $||D_n||^2$ we use Eq.~\eqref{eq:SYK_ensemble_avg_conventions} in 
\be 
\begin{split}
\sum_{| I_{j} \cap I_l | = 0} \sum_{| K_{j} \cap K_l | = 0} \expval{J_{I_1} J_{I_2} \ldots J_{I_n} J_{K_1} J_{K_2} \ldots J_{K_n}} \\
\tr(\Psi_{I_1} \Psi_{I_2} \ldots \Psi_{I_n}  \Psi_{K_1} \Psi_{K_2} \ldots \Psi_{K_n} ),   
\end{split}
\ee
The ensemble average over coupling can be interpreted as a microscopic chord contraction within a fixed size sector. Only those configurations contribute where each $K_i$ is paired up with some $I_j$. This holds even if the couplings $J_{I}$ are not Gaussian distributed, as long as they are independent with zero mean. Unlike the usual chord problem, there are no factors of $e^{-\lambda}$, because we explicitly projected onto the sector where all $\Psi_{I}$ commute. There are $n!$ pairings between the bra and the ket, from which 
\be 
\begin{split}
   ||D_n||^2 &= n! \Big( \frac{\sJ^2}{\lambda} \Big)^n \frac{N!}{(N-n q)! (q!)^n} \binom{N}{q}^{-n}, \\
  &=  n! \Big( \frac{\sJ^2}{\lambda} \Big)^n \big( 1 - \frac{n(n-1) q^2}{2N} + O(N^{-2})\big).
\end{split}
\label{eq:syk_series_dn} \ee
$r_n$ can be computed using 
\be \begin{split}
\mu_4 &= \mu^2_2 \Big[ 2 + \binom{N}{q}^{-1} \sum^{j=q}_{j=0}  (-1)^j \binom{q}{j} \binom{N-q}{q-j}  \Big] \\ 
&= \mu^2_2 \bigg(3 -  \frac{2q^2}{N} + O(N^{-2} ) \bigg), 
\end{split} \ee 
from which it follows that 
\be r_n = n! \mu^2_2 \big( 1- \frac{n(n-1) q^2}{2 N} + O(N^{-2} ) \big), \ee 
and consequently $\epsilon_n = O(N^{-2})$. This result assumes ensemble averaging of the denominator $r_n$ and the numerator $||D_n ||^2$ separately instead of ensemble averaging $\frac{||D_n ||^2}{r_n}$. This does not affect the credibility of this result, since we are inspecting the $1/N$ correction and because these sums have a statistical fluctuations of $O(N^{-q})$ in SYK.

There is a related microscopic argument for SYK which strongly suggests that the $\frac{1}{N^2}$ suppression holds for each disorder realization. The probability of the pair of strings $\Psi_{I}, \Psi_{J}$ having $j$ intersections is given by $\binom{q}{j} \binom{N-q}{q-j}  \binom{N}{q}^{-1}$. If there is exactly one intersection in the expansion of $P_n(H)$ in Majorana strings, then that contribution gets canceled in $J_{I_1} J_{I_{2}} ( \Psi_{I_1}  \Psi_{I_2} + \Psi_{I_2}  \Psi_{I_1} )$ because $\Psi_{I_2}  \Psi_{I_1} = -\Psi_{I_1} \Psi_{I_2}$. Pairwise intersections of $k \geq 2$ fermions occur with probability $O(N^{-k})$. For $\ket{2}$, the size-$2q-2$ term vanishes and the size-$2q-4$ term carries a coefficient suppressed by $\frac{1}{N^2}$. We conjecture that after orthogonalizing against $\ket{m}$  with $m < n$, a similar mechanism suppresses the size $nq-4$ projection of $\ket{n}$ by $O(N^{-2})$ for each disorder realization. Variants of SYK with commuting chords have been studied in \cite{gao2024_commuting_syk, berkooz2024_commuting, Almheiri_2024}. 
 
\section{Size of Complexified Time Evolution Operator \label{sec:size_of_tfd}}
We denote the normalized form of the state $\ket{g(H)}$, where $g(H)=e^{-z H}$ for $z \in \mathbb C$ by
\be \ket{z} :=\frac{1}{\sqrt{Z(2 \Re(z))}} e^{-z H} \ket{\Omega}. \ee 
It reduces to the thermofield double for real $z= \frac{1}{2}\beta$. The mean size is given by 
\be \frac{1}{Z(2 \Re(z) )} \sum_{i} \tr( e^{-z^{*} H } \mS_i(e^{-z H} ) ) \label{eq:size_superop_tfd} \ee 
where we used the state operator map to write Eq.~\eqref{eq:size_superop_tfd} in terms of the size superoperator 
\be \mathcal{S}_i(O) = \frac{3}{4} O - \frac{1}{4} \bigg(\sigma^{x}_i O \sigma^{x}_i +\sigma^{y}_i O \sigma^{y}_i +\sigma^{z}_i O \sigma^{z}_i  \bigg). \ee
Hence Eq.~\eqref{eq:size_superop_tfd} evaluates to the sum of two point functions 
\be \frac{1}{4} \big( 3 N -  \sum_{\alpha} C_{\alpha}(z) \big), \label{eq:size_tfd_2pt_function}  \ee
in terms of the site-averaged correlation function
\be C_{\alpha}(z) = \frac{1}{Z(2 \Re(z) )} \sum_{i} \tr(e^{-z^{*} H }  \sigma^{\alpha}_i e^{-z H } \sigma^{\alpha}_i), \ee 
evaluated at complex time $z$. For $O(1)$ value of $\abs{z}$, we expect each spin autocorrelation to decay by a finite amount and hence Eq.~\eqref{eq:size_tfd_2pt_function} evaluates to an $O(N)$ number. This is supported by the leading order term in a small-$\abs{z}$ expansion:
\be \bra{z} S \ket{z} = q \abs{z}^2 \tr(H^2) + O(\abs{z}^3), \label{eq:size_taylor_distribution}\ee
which shows that the extensive scaling of $\tr(H^2)$ implies the extensive scaling of $\bra{z}S\ket{z}$. Since the Krylov state $\ket{n}$ does not have a size component larger than $nq$, it follows that in $\ket{z} = \sum_n c^{z}_n \ket{n}$ the coefficients $\abs{c^z_n}^2$ must have support on $n > \frac{1}{q} \bra{z} S \ket{z}$. A similar observation was used to derive the bound $\sum_n \abs{c^z_n}^2 n \geq \bra{z} S \ket{z} $ in \cite{Parker_2019}. The expansion for complexity is also straightforward:
\be \sum_n \abs{c^z_n}^2 n = \abs{z}^2 \tr(H^2) + O(\abs{z}^3), \ee 
and consistent with our identification, for sufficiently small $\abs{z}$.
\end{document}